\documentclass[epj]{svjour}
\usepackage{graphics}
\usepackage{amsmath}

\newcommand{\apj}{Astrophys. J.}
\newcommand{\apjl}{Astrophys. J. Lett.}
\newcommand{\prd}{Phys. Rev. D}
\newcommand{\prl}{Phys. Rev. Lett.}
\newcommand{\prc}{Phys. Rev. C}
\newcommand{\araa}{Annual Review of Astronomy and Astrophysics}
\newcommand{\nat}{Nature}
\newcommand{\mnras}{Mon. Not. R. Astron. Soc.}

\begin{document}
\title{Exploring properties of high-density matter through remnants of neutron-star mergers}

\author{Andreas Bauswein\inst{1,2} \and Nikolaos Stergioulas\inst{1} \and Hans-Thomas Janka\inst{3}
}                     
\offprints{}          
\institute{Department of Physics, Aristotle University of
  Thessaloniki, GR-54124 Thessaloniki, Greece \and Heidelberger Institut f\"ur Theoretische Studien, Schloss-Wolfsbrunnenweg~35, D-69118~Heidelberg, Germany \and Max-Planck-Institut f\"ur
  Astrophysik, Karl-Schwarzschild-Str.~1, D-85748~Garching, Germany}
\date{Received: date / Revised version: date}
%
\abstract{Remnants of neutron-star mergers are essentially massive, hot, differentially rotating neutron stars, which are initially strongly oscillating. As such they represent a unique probe for high-density matter because the oscillations are detectable via gravitational-wave measurements and are strongly dependent on the equation of state. The impact of the equation of state for instance is  apparent in the frequency of the dominant oscillation mode of the remnant. For a fixed total binary mass a tight relation between the dominant postmerger oscillation frequency and the radii of nonrotating neutron stars exists. Inferring observationally the dominant postmerger frequency thus determines neutron star radii with high accuracy of the order of a few hundred meters. By considering symmetric and asymmetric binaries of the same chirp mass, we show that the knowledge of the binary mass ratio is not critical for this kind of radius measurements. We perform simulations which 
show that initial intrinsic neutron star rotation is unlikely to affect this method of constraining the high-density equation of state. We also summarize different possibilities about how the postmerger gravitational-wave emission can be employed to deduce the maximum mass of nonrotating neutron stars. We clarify the nature of the three most prominent features of the postmerger gravitational-wave spectrum and argue that the merger remnant can be considered to be a single, isolated, self-gravitating object that can be described by concepts of asteroseismology. We sketch how the consideration of the strength of secondary gravitational-wave peaks leads to a classification scheme of the gravitational-wave emission and postmerger dynamics. The understanding of the different mechanisms shaping the gravitational-wave signal yields a physically motivated analytic model of the gravitational-wave emission, which may form the basis for template-based gravitational-wave data analysis. We explore the observational 
consequences 
of a scenario of two families of compact stars including hadronic and quark matter stars. We find that this scenario leaves a distinctive imprint on the postmerger gravitational-wave signal. In particular, a strong discontinuity in the dominant postmerger frequency as function of the total mass will be a strong indication for two families of compact stars.
} 
\maketitle
\section{Introduction}
\label{intro}

By means of intense experimental efforts gravitational waves (GWs) are expected to be detected within the next years. To date, their existence is only indirectly proven by the observations of neutron star (NS) binaries~\cite{1989ApJ...345..434T,2003LRR.....6....5S,2010ApJ...722.1030W}, in particular the Hulse-Taylor binary pulsar PSR B1913+16~\cite{1975ApJ...195L..51H}. The orbital period and orbital separation of this binary decrease precisely as determined by General Relativity, which predicts that the GW emission continuously reduces the system's angular momentum and energy. As the orbital frequency increases with time the GW emission becomes stronger, and thus the decay of the orbit proceeds increasingly faster. For PSR B1913+16 the orbital period will decrease from currently 7.75~hours to a few milliseconds in the next $3\times 10^8$~years. The corresponding GW emission reflects the dynamics of the orbital motion and thus increases in frequency and amplitude resulting in a chirp-like signal, 
which is determined by the binary masses (see e.g. the first milliseconds in Fig.~\ref{fig:toysignal} or Figs.~5 and~6 in~\cite{2012PhRvD..86f3001B}). Only during the last seconds of this so-called inspiral (named after the shape of the stellar trajectories) the signal will enter the sensitivity band of the existing and upcoming GW detectors between roughly 10 and 10000~Hz~\cite{2015CQGra..32b4001A,2013PhRvD..88d3007A,2015CQGra..32g4001T}. Importantly, the total population of NS binaries within a distance of a few 100~Mpc to our Galaxy (i.e. the range that will be covered by GW instruments) is expected to be sufficiently high such that per year of the order of 40 binaries in the last phase of their inspiral will enter the sensitivity band of GW detectors~\cite{2010CQGra..27q3001A}.

The continuously decreasing orbital separation of NS binaries inevitably leads to the merging of the binary components. On a dynamical time scale of the order of a millisecond the two stars form a single, massive, differentially rotating object (see e.g. Fig.~\ref{fig:lapsesnap} and Fig.~2 in~\cite{2015arXiv150203176B}). Mostly because of its rotation the merger remnant can be supported against the gravitational collapse even if the total binary mass significantly exceeds the maximum mass of non-rotating NSs. Only for very high total binary masses a prompt gravitational collapse occurs. The stellar object forming during a coalescence of two NSs is very interesting from the perspective of high-density matter physics\footnote{Within this paper the term ``high density'' refers to densities roughly above nuclear saturation density, which is the regime most relevant for the stellar structure of NSs.} because it resembles a massive, hot, rotating NS, which is strongly oscillating (see 
e.g.~\cite{2010CQGra..27k4002D,2011GReGr..43..409A,BaumgarteShapiro,2012LRR....15....8F,Rezzolla} for reviews). The oscillations are induced by the merging process, notably the fundamental quadrupolar fluid mode of the remnant is strongly excited and dominates the postmerger GW signal. The frequencies of the different excited oscillation modes are characteristic of the remnant's mass and of the high-density equation of state (EoS). The mass of the remnant is approximately identical to the total binary mass. The binary mass as well as (in principle) most of the oscillation frequencies of the merger remnant are observationally accessible by GWs. Therefore, NS mergers offer the possibility of NS asteroseismology and of inferring properties of the high-density EoS, or 
equivalently, of stellar properties of non-rotating NSs, via observations with the upcoming and existing GW detectors. The merger rate is estimated to be roughly of the order of 
$10^{-5}$ events per year per Milky way equivalent galaxy, which corresponds to a detection rate of about 40 inspiral detections per 
year for Advanced LIGO and Advanced Virgo when the instruments operate at their design sensitivity (see e.g.~\cite{2010CQGra..27q3001A} for a collection of rate estimates and their uncertainties). With these instruments the oscillations during the postmerger phase are detectable for relatively nearby events~\cite{2014PhRvD..90f2004C,Clark2015}, which is an exciting prospect given that already a single event is sufficient to provide tight constraints on the EoS~\cite{2012PhRvL.108a1101B,2012PhRvD..86f3001B}.

Because of its high mass, the postmerger remnant is particularly interesting since the central densities are relatively high compared to those of many observed NSs in binary systems, which have masses in the range of 1.2~$M_\odot$ to 1.5~$M_\odot$ (see e.g.~\cite{2012ARNPS..62..485L}). This implies that the GW emission of the postmerger phase probes a density regime that is hardly accessible by other observational methods. In particular, attempts to infer NS properties via finite-size effects in the GW signal during the inspiral phase preceeding the merger are restricted to the mass range of the individual components of the binary (see e.g.~\cite{2010PhRvD..81l3016H,2010PhRvD..81h4016D,2012PhRvD..85l3007D,2013arXiv1310.8288F,2013arXiv1306.4065R,2013PhRvL.111g1101D,2014PhRvD..89j3012W,2015arXiv150305405A,2015PhRvD..91d3002L,2015arXiv150802062C}). In this sense the postmerger oscillations represent a complementary approach by providing insights to NS properties at higher masses and higher densities (e.g.~\cite{
2014PhRvD..90b3002B,BausweinWroclaw}). 
Moreover, the collapse behavior of NS merger remnants may reveal the properties of the very high density regime, which is decisive for determining the threshold of the gravitational collapse~\cite{2013PhRvL.111m1101B}. The consideration of 
the collapse behavior may be particularly rewarding for establishing limits on the maximum mass of nonrotating NSs, which is challenging given the paucity of NS systems with masses close to the collapse threshold, while the mass range of merger remnants partially overlaps with the mass range where the gravitational collapse is expected to take place for various EoSs. Therefore, an upper limit on the maximum mass may be established through NS mergers, which may be hard to derive through other types of observations.

The basic impact of the EoS on the merger dynamics and GW signal can be described as follows. Stiff EoSs lead to NSs with large radii. Such stars can be deformed more easily under the influence of an external tidal field. Consequently, finite size effects during the inspiral set in at a larger orbital separation, i.e. a lower orbital frequency, and the stars finally merge at a relatively low orbital frequency. In contrast, soft EoSs yield more compact NSs, which in comparison to larger NSs behave more like point particles during the late inspiral phase. Stars described by soft EoSs are harder to deform and during the inspiral they reach higher orbital velocities before they merge. This also implies a higher linear velocity before merging and results in a merger with a higher impact velocity. The stiffness also affects the dynamics of the postmerger phase and, in particular, the frequencies of the excited oscillation modes. A stiff EoS leads to a relatively large merger remnant, whose quadrupolar fluid 
oscillation frequency is relatively low, since it scales approximately with the mean density. In the case of a soft EoS, the merger remnant is more compact and thus oscillates at higher frequencies. (For such EoSs the higher impact velocity during merging additionally leads to a stronger excitation of the quasi-radial oscillation mode of the remnant.)

In this paper we summarize the current status of possibilities to deduce NS properties and EoS constraints from GW observations of the postmerger phase of NS mergers. We focus mostly on the most prominent oscillation mode in the GW spectrum of the postmerger phase, the fundamental quadrupolar fluid mode, since this is the feature which is most likely to be detected. We prepend a discussion of mass measurements through GW chirp signals of NS mergers, which is essential to NS radius measurements by the observation of GWs from the postmerger phase as well as for constraints on the maximum mass of nonrotating NSs. We supplement a review of previous findings by new results, which for instance clarify the role of intrinsic NS rotation of the binary components. Also, we provide further evidence that the merger remnant can be considered to be a single, self-gravitating object and can thus be investigated by concepts of asteroseismology~\cite{2011MNRAS.418..427S}. By considering certain oscillation modes and 
secondary features of the postmerger 
GW 
spectrum, we 
arrive at a unified classification scheme of the postmerger dynamics and GW emission. We also present an analytic model of the GW emission, which may represent a step towards the construction of GW templates that can be employed in GW data analysis. Along the lines of the current Topical Issue of ``Exotic Matter in Neutron Stars'' we also discuss a scenario of the existence of two families of compact stars~\cite{2014PhRvD..89d3014D} and identify certain observational features which can test these ideas. This discussion represents an example of how the postmerger phase can be employed to test high-density matter properties at densities which are not accessible through NSs of masses below $\sim 1.5~M_\odot$.

The results presented in this paper rely mostly on the calculations discussed in~\cite{2012PhRvL.108a1101B,2012PhRvD..86f3001B,2013PhRvL.111m1101B,2014PhRvD..90b3002B,2015arXiv150203176B}, where further information can be found. Additional new findings laid out here are obtained within the same physical and numerical model, for which further details are provided in~\cite{2002PhRvD..65j3005O,2007A&A...467..395O,2010PhRvD..82h4043B,2010PhRvD..81b4012B,2012PhRvD..86f3001B}. We only consider mergers from quasi-circular orbits (for a discussion of tidal capture events see e.g.~\cite{2010ApJ...720..953L}). This paper is written in a modular way such that the individual sections are mostly self-contained and the reader can skip certain topics. Introductory remarks on mass measurements via GWs are presented in Sect.~\ref{sec:mass}. NS radius determinations via the postmerger phase are discussed in Sect.~\ref{sec:rad}. We address the collapse behavior of NS merger remnants and outline possibilities to measure the 
maximum mass of nonrotating NSs in Sect.~\ref{sec:col}. The origin of the major and of secondary features of the GW spectrum are 
explained in Sect.~\ref{sec:sec} as well as the dependencies of the secondary GW peaks, which lead to a classification scheme of the postmerger dynamics and GW emission. In Sect.~\ref{sec:toy} we provide details on 
our analytic model of the postmerger GW emission. The impact of intrinsic NS rotation is considered in Sect.~\ref{sec:rot}. The possible observational implications of a scenario of two distinct families of compact stars is explored in Sect.~\ref{sec:two}. We summarize and conclude in Sect.~\ref{sec:sum}.

\section{Mass measurements}\label{sec:mass}
The strongest part of the GW signal of a NS merger originates from the preceding inspiral phase, where the system continuously losses angular momentum and energy by the emission of GWs. The losses lead to  inspiralling trajectories of the binary components, while the GW signal is determined by the orbital motion of the NSs. The inspiral accelerates, which results in a chirp-like GW signal with an increasing amplitude and an increasing frequency. Finite-size effects become important only for the last orbits before merging and may yield information on the NS EoS, specifically on the tidal deformability, which for fixed mass approximately scales with the stellar radius (see e.g.~\cite{2010PhRvD..81l3016H,2010PhRvD..81h4016D,2012PhRvD..85l3007D,2013arXiv1310.8288F,2013arXiv1306.4065R,2013PhRvL.111g1101D,2014PhRvD..89j3012W,2015arXiv150305405A,2015PhRvD..91d3002L,2015arXiv150802062C}).
\begin{figure}

\resizebox{0.5\textwidth}{!}{%
  \includegraphics{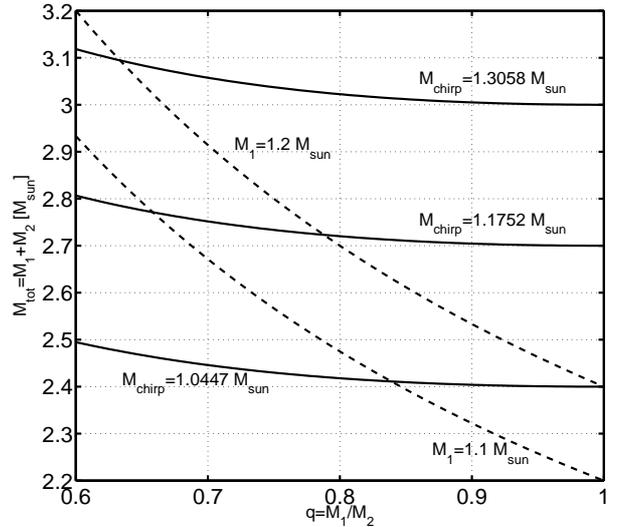}
}
\caption{Total binary mass $M_\mathrm{tot}=M_1+M_2$ as a function of the mass ratio $q=M_1/M_2$ for systems with constant chirp mass (solid lines), which will be measured with high precision by GW detectors. Dashed lines show systems where the less massive component has a mass of $M_1=1.1~M_\odot$ and $M_1=1.2~M_\odot$, which represent reasonable lower limits on the NS mass, and thus restrict the allowed systems to the right of the respective dashed curves.}
\label{fig:mchirp}
\end{figure}

The key parameter determining the GW signal during the inspiral phase is the so-called chirp mass apart from other parameters like the distance to the source, the inclination of the binary, the orientation of the instruments and so on. The chirp mass is given by
\begin{equation}\label{eq:chirp}
M_\mathrm{chirp} = (M_1 M_2)^{3/5} (M_1+M_2)^{-1/5}
\end{equation}
with the NS masses $M_1$ and $M_2$. Being the crucial quantity to determine the signal, the chirp mass will be measured with very high accuracy since it is one of the parameters used to construct the template bank for matched filtering searches that will be employed to actually detect GWs from binaries~\cite{1993PhRvD..47.2198F,1994PhRvD..49.2658C,1996CQGra..13.1279J,2005PhRvD..71h4008A,2008CQGra..25r4011V,2012PhRvD..85j4045V,2013ApJ...766L..14H,2013PhRvD..88f2001A,2014ApJ...784..119R,2015PhRvD..91d2003V,2015arXiv150805336F}. In contrast, the mass ratio can only be measured with higher accuracy for nearby events~\cite{1993PhRvD..47.2198F,1994PhRvD..49.2658C,1996CQGra..13.1279J,2005PhRvD..71h4008A,2008CQGra..25r4011V,2012PhRvD..85j4045V,2013ApJ...766L..14H,2013PhRvD..88f2001A,2014ApJ...784..119R,2015PhRvD..91d2003V,2015arXiv150805336F}.

It can be seen from Eq.~\eqref{eq:chirp} that a measurement of the chirp mass alone does not suffice to determine the individual binary masses, which can only be achieved by determining the mass ratio  with sufficient precision. However, for NS binaries the chirp mass alone provides already a very good estimate of the total binary mass $M_\mathrm{tot}=M_1+M_2$. Figure~\ref{fig:mchirp} shows the total binary mass as a function of the mass ratio $q=M_1/M_2\leq 1$ for sequences of constant chirp mass (solid lines). For a constant chirp mass the total binary mass depends only weakly on the mass ratio. The dashed lines define the binary systems, where the less massive component has a gravitational mass of 1.1~$M_\odot$ and 1.2~$M_\odot$, respectively. Under the reasonable assumption that NSs cannot be less massive than 1.2~$M_\odot$ (see discussion below), a measured chirp mass restricts the possible binary parameters to the right of the dashed curve for $M_1=1.2~M_\odot$. This implies that for instance a 
measured chirp mass 
of 1.1752~$M_\mathrm{tot}$ restricts the total binary mass of this event to the range $2.7~M_\odot \leq M_\mathrm{tot} \leq 2.73~M_\odot$ (without any information about the mass ratio). From observed NS binaries (see e.g. listing in~\cite{2012ARNPS..62..485L}) and theoretical population synthesis 
studies (e.g.~\cite{2012ApJ...759...52D}) it is expected that 
most binaries have a chirp mass close to 1.1752~$M_\mathrm{tot}$ (corresponding to a total binary mass of roughly 2.7~$M_\odot$). This means that for the most likely merger event the total binary mass can be determined very accurately from the chirp mass measurement only. Even for very massive binaries (with a chirp mass of 1.3058~$M_\odot$) the total binary mass can be inferred with a precision of $\pm 0.05~M_\odot$.

To support the arguments above we remark that recent core-collapse simulations for a large number of progenitor stars find a minium NS rest mass of about 1.3~$M_\odot$~\cite{2015arXiv150307522E}. A NS rest mass of 1.3~$M_\odot$ corresponds to a gravitational mass of roughly 1.2~$M_\odot$, somewhat dependent on the EoS. These findings justify to assume that NS binaries are unlikely to host NSs less massive than 1.2~$M_\odot$. Also other formation channels like an accretion-induced collapse of a white dwarf are unlikely to form less massive NSs. See also the discussion of the minimum NS mass in~\cite{2012ARNPS..62..485L}. Finally, we note that a white dwarf-NS binary might have a chirp mass in the range which is typical of NS-NS binaries, and a more extreme mass ratio if the white dwarf is less massive than 1.2~$M_\odot$. The merger of a white dwarf-NS binary will, however, lead to significantly altered inspiral dynamics~\cite{2011PhRvD..84j4032P} and thus a different GW signal prior to merging such that a 
white dwarf-NS merger can be easily distinguished from a NS binary merger.

The considerations above play a role for the following discussions, because methods to infer EoS properties from NS mergers rely on the ability to measure at least the total binary mass from the inspiral GW signal. Information on the mass ratio and thus the individual binary component masses is not critical, but improves the constraints on NS properties. As outlined above, the determination of the total binary mass via the chirp mass alone represents the absolute minimum of what is achievable with existing and upcoming GW detectors. However, it is important to stress that for distances of order 50~Mpc, which allow to deduce EoS properties, the individual binary masses are expected to be recovered with a precision of a few per cent~\cite{1993PhRvD..47.2198F,1994PhRvD..49.2658C,1996CQGra..13.1279J,2005PhRvD..71h4008A,2008CQGra..25r4011V,2012PhRvD..85j4045V,2013ApJ...766L..14H,2013PhRvD..88f2001A,2014ApJ...784..119R,2015PhRvD..91d2003V,2015arXiv150805336F}. For instance, in~\cite{2014ApJ...784..119R} the 
individual NS masses have 
been recovered within about 10 per cent at a distance of 100~Mpc. Assuming that the error scales linearly with the distance, an accuracy of a few per cent could be reached in determining the individual masses of the merging NSs at a distance of a $\sim 50$~Mpc. Therefore, we will work in the following under the condition that the individual masses can be measured sufficiently accurately for our purposes and that, in particular, the total binary mass can be determined very well.

\section{Radius measurements}\label{sec:rad}
The outcome of a NS merger essentially depends on the EoS of high-density matter and the total binary mass. The binary mass ratio has a modest impact, while the initial NS rotation and magnetic fields have only a small influence.) Simulations show that for binary masses of about 2.7~$M_\odot$ the merging leads to the formation of a massive, hot, differentially rotating NS remnant for most EoSs, even for EoS models which yield a maximum mass of nonrotating NSs significantly below 2.7~$M_\odot$, e.g.~\cite{1994PhRvD..50.6247Z,1996A&A...311..532R,2000ApJ...528L..29B,2003ApJ...583..410L,2005PhRvL..94t1101S,2005PhRvD..71h4021S,2006PhRvD..73f4027S,2007A&A...467..395O,2007PhRvL..99l1102O,2008PhRvD..77b4006A,2008PhRvD..78b4012L,2008PhRvD..78h4033B,2009PhRvD..80f4037K,2011MNRAS.418..427S,2011PhRvD..83d4014G,2011PhRvD..83l4008H,2011PhRvL.107e1102S,2012PhRvL.108a1101B,2012PhRvD..86f3001B,2012PhRvD..86f4032P,2013MNRAS.430.2585R,2013PhRvL.111m1101B,2013PhRvD..88d4026H,2014PhRvD..89j4021B,2014PhRvD..90b3002B,2014PhRvL.113i1104T,2014PhRvD..90d1502K,2014MNRAS.437L..46R,2014arXiv1412.3240T,2014arXiv1411.7975K,2015arXiv150203176B,2015arXiv150401764B,2015arXiv150501607P,2015arXiv150707100D,2015PhRvD..92h4064D,2015arXiv150908804D,2015arXiv151006398F}. The rapid rotation stabilizes the remnant against gravitational collapse. According to pulsar observations in NS binaries most systems are expected to have a total mass of roughly 2.7~$M_\odot$ (see e.g. the compilation of binary masses in~\cite{2012ARNPS..62..485L}), which is also supported by theoretical studies of the binary population, e.g.~\cite{2012ApJ...759...52D}. Given this mass range, the most likely outcome of a merger event should be expected to be the formation of a NS remnant (see~\cite{2013PhRvL.111m1101B} for the EoS dependence of collapse behavior).
\begin{figure}

\resizebox{0.5\textwidth}{!}{%
  \includegraphics{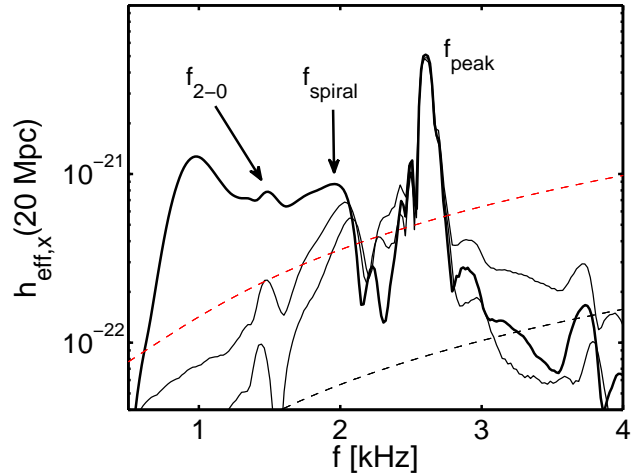}
}
\caption{GW spectrum of the cross polarization of a 1.35-1.35~$M_\odot$ merger with the DD2 EoS~\cite{2010NuPhA.837..210H,2010PhRvC..81a5803T} along the polar direction at a distance of 20~Mpc. $h_\mathrm{eff}=\tilde{h}(f)\cdot f$ with the Fourier transform of the waveform $h_{\times}$ and frequency $f$. $f_\mathrm{peak}$, $f_\mathrm{spiral}$ and $f_{2-0}$ are particular features of the postmerger phase, which can be associated with certain dynamical effects in the remnant. Since the simulation started only a few orbits before merging, i.e. at a relatively high orbital frequency, the power at lower frequencies (below $\sim$1~kHz) is massively underrepresented in the shown spectrum, and the low-frequency part of the spectrum does not show the theoretically expected power-law decay. The thin solid lines display the spectra of the GW signal of the postmerger phase only revealing that the peaks are indeed generated in the postmerger phase. Dashed lines show the expected unity SNR sensitivity curves of 
Advanced LIGO~\cite{2010CQGra..27h4006H} (red) and of the Einstein Telescope~\cite{2010CQGra..27a5003H} (black).}
\label{fig:spect}
\end{figure}

A typical GW spectrum of a NS merger resulting in the formation of a NS remnant is shown in Fig.~\ref{fig:spect}. The spectrum is computed from a simulation of a 1.35-1.35~$M_\odot$ merger with the DD2 EoS~\cite{2010NuPhA.837..210H,2010PhRvC..81a5803T}. The dominant GW oscillation frequency $f_\mathrm{peak}$ of the postmerger phase is clearly visible as a pronounced peak in the kHz range, which is produced by the dominant quadrupolar remnant oscillation. Apart from the main peak one recognizes several additional peaks, whose nature will be discussed in Section~\ref{sec:sec}. By computing the spectrum of the GW signal of the postmerger phase only, one can show that the different features are related to the postmerger phase (see thin lines in Fig.~\ref{fig:spect}). 

\begin{figure}

\resizebox{0.5\textwidth}{!}{%
  \includegraphics{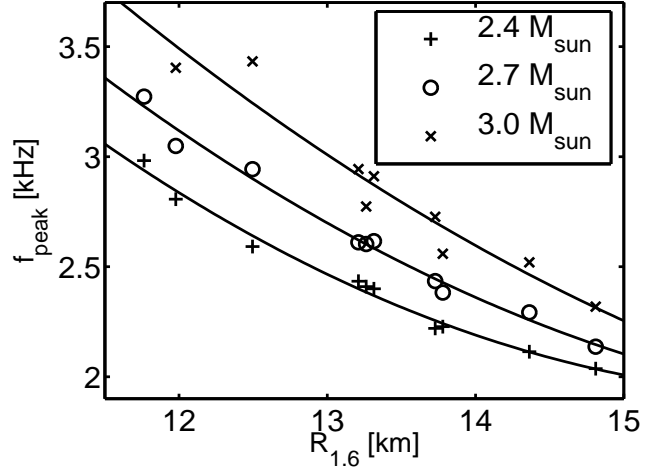}
}
\caption{Dominant postmerger GW frequency $f_\mathrm{peak}$ as function of the radius $R_{1.6}$ of a nonrotating NS with a gravitational mass of 1.6~$M_{\odot}$ for different EoSs and different total binary masses (plus signs for 2.4~$M_{\odot}$, circles for 2.7~$M_{\odot}$, crosses for 3.0~$M_{\odot}$) and a mass ratio of unity. The solid lines are least-square fits to the data of the different binary masses.}
\label{fig:fpeak}
\end{figure}

The dominant oscillation frequency $f_\mathrm{peak}$ depends in a specific way on the high-density EoS~\cite{2012PhRvL.108a1101B,2012PhRvD..86f3001B}. This is understandable since the EoS affects the size of the remnant, which in turn determines its oscillation frequency (see Fig.~13 in~\cite{2012PhRvD..86f3001B}). The structure of the remnant is also influenced by its angular momentum. The available angular momentum, however, is given by the dynamics of the late inspiral/merging phase, which is fully determined by the stellar structure of the inspiralling stars and thus also depends on the EoS in a particular way. The strong EoS dependence of the peak frequency can be expressed as follows. An EoS which is used in a given simulation, can be conveniently characterized by the radii of nonrotating NSs, which are uniquely determined by this EoS through the stellar structure equations (Tolman-Oppenheimer-Volkoff equations~\cite{1939PhRv...55..364T,1939PhRv...55..374O}). Specifically, for a set of calculations 
with 
a fixed total binary mass but different EoSs\footnote{Except for some models considered in Sect.~\ref{sec:col}, the EoSs discussed in this study are temperature dependent and include electrons, positrons and photons, while neutrino contributions are neglected. With regard to the resulting stellar properties these EoSs cover a representative range, which, for example, can be seen from the range of radii in Fig.~\ref{fig:fpeak} and maximum masses in Fig.~\ref{fig:mmaxfpeak}.}, we relate the peak frequency, which is extracted from a 
simulation with a given EoS, to the radius of a nonrotating NS (described by the same EoS) with a fixed fiducial mass. A natural choice is to employ the NS radius for a mass of $M_{\mathrm{NS}}=M_\mathrm{tot}/2$, which for symmetric binaries is just the radius of the inspiralling NSs (more precisely, at infinite orbital separation). In this case a clear correlation is found, where EoSs leading to more compact NSs yield higher postmerger GW frequencies (see Fig.~12 in~\cite{2012PhRvD..86f3001B}, which shows this relation for $M_\mathrm{tot}=2.7~M_\odot$). (Alternatively, one can use the compactness $C=G M_\mathrm{NS}/(c^2 R(M_\mathrm{NS}))$ of fiducial nonrotating NS models, which is equivalent to employing the radius $R(M_\mathrm{NS})$.) 

The empirical relation between $f_\mathrm{peak}$ and $R(M_\mathrm{NS})$ is very tight, which implies that a measurement of the peak frequency can be used to determine the unknown radius of a nonrotating NS with a fixed mass by simply inverting the empirical relation~\cite{2012PhRvL.108a1101B,2012PhRvD..86f3001B}. Thus, a future detection of the GW postmerger phase and extraction of the peak frequency (see~\cite{2014PhRvD..90f2004C,Clark2015}) will yield strong constraints on the high-density EoSs. In~\cite{2012PhRvL.108a1101B,2012PhRvD..86f3001B} the largest deviation of the empirical data from a fit is only a few hundred meters. The accuracy of a radius determination by the postmerger GW signal is mostly affected by two sources of error. One error is the uncertainty of the measurement of the peak frequency. Apart from this, one should take into account deviations between the data and the fit to the data describing the empirical relation. A measurement of the peak frequency (of the true EoS) does not reveal 
in which 
way the 
measured 
frequency slightly deviates from the empirical relation. Hence, one conservatively has to assume that the true data point may deviate as much as the largest deviation found in the large sample of candidate EoSs. 

\begin{figure}

\resizebox{0.5\textwidth}{!}{%
  \includegraphics{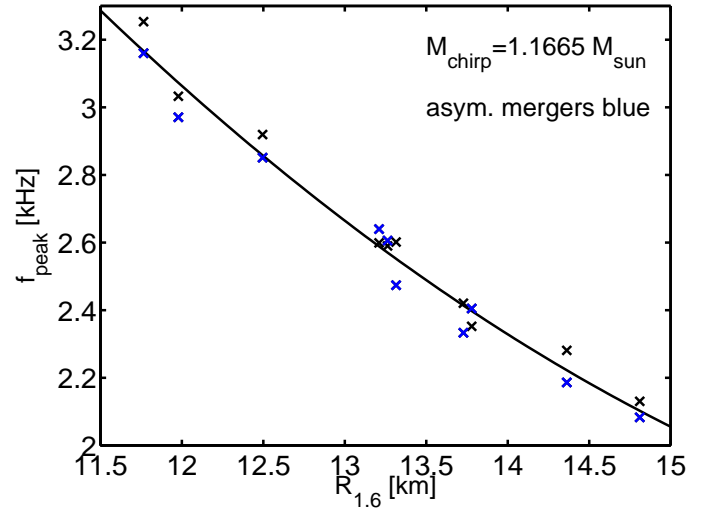}%
}
\caption{Dominant postmerger GW frequency $f_\mathrm{peak}$ as function of the radius $R_{1.6}$ of a nonrotating NS with a gravitational mass of 1.6~$M_{\odot}$ for symmetric and asymmetric binaries with a chirp mass of $M_\mathrm{chirp}=1.1665~M_\odot$ for different EoSs. Black symbols display data for 1.34-1.34~$M_\odot$ mergers, while blue symbols exhibit peak frequencies of 1.2-1.5~$M_\odot$ mergers. The solid line is a least-square fit to the data of the different binary masses.}
\label{fig:asym}
\end{figure}

The peak frequency has been shown to be measurable with very high precision by a coherent burst search analysis~\cite{2014PhRvD..90f2004C}. In this study waveforms from numerical models were superimposed with the recorded data stream of previous GW detector science runs, which simulates the noise of the future instruments. The model waveforms were injected at random times and the noise was rescaled to the anticipated sensitivity of the second-generation GW detectors Advanced LIGO and Virgo. The existing GW data analysis pipeline was able to recover the injected signal and to determine the peak frequency with an accuracy of $\sim 10$~Hz, which is smaller than the spread in the empirical relation between $f_\mathrm{peak}$ and the NS radius. This implies that the radii of the inspiralling stars can be determined with a precision of a few hundred meters.

These considerations show that the larger contribution to the error of a radius measurement originates from the scatter in the empirical relation between $f_\mathrm{peak}$ and $R(M_\mathrm{NS})$. In this context, the following observation is important. One has the freedom to choose any fiducial NS mass different from $M_{\mathrm{NS}}=M_\mathrm{tot}/2$ for characterizing a given EoS by the TOV radius $R(M_\mathrm{NS})$. Empirically, it turns out that using a fiducial NS mass somewhat larger than $M_{\mathrm{NS}}=M_\mathrm{tot}/2$ leads to tighter relations between $f_\mathrm{peak}$ and $R(M_\mathrm{NS})$. This is exemplified in Fig.~\ref{fig:fpeak}. For $M_\mathrm{tot}=2.7~M_\odot$ (circles in Fig.~\ref{fig:fpeak}) the maximum deviation between the data and a fit amounts to only $\sim 175$~meters if $M_\mathrm{NS}=1.6~M_\odot$ is chosen. This implies that the measurement of the dominant postmerger frequency for $M_\mathrm{tot}=2.7~M_\odot$ determines the radius of a nonrotating 1.6~$M_\odot$ NS 
with an accuracy of better than 200~meters. 

It is natural that a fiducial mass of $M_\mathrm{NS}=1.6~M_\odot$ is somewhat more appropriate than $M_\mathrm{NS}=1.35~M_\odot$ for characterizing the postmerger oscillations of 1.35-1.35~$M_\odot$ mergers ( $M_\mathrm{tot}=2.7~M_\odot$). The maximum densities in the massive, rotating merger remnant are higher than in the initial NSs and they are comparable to the central densities of nonrotating, static NSs with a mass of roughly 1.6~$M_{\odot}$ (see e.g. Fig.~15 in~\cite{2012PhRvD..86f3001B}). For this reason, nonrotating NSs with $M_{\mathrm{NS}}>M_\mathrm{tot}/2$ better represent the density regime encountered in the merger remnant and thus provide a better description of the EoS.

Figure~\ref{fig:fpeak} also shows that similar empirical relations hold for other binary masses. The different symbols display the peak frequencies from calculations with total binary masses of 2.4~$M_{\odot}$ (plus symbol), 2.7~$M_{\odot}$ (circles), and 3.0~$M_{\odot}$ (crosses) for symmetric mergers. The solid lines are least-square fits to the data. One recognizes that also for other binary masses there are only small deviations from the empirical relations on the order of only a few 100 meters. Hence, these relations can be used for determining NS radii with similar precision after the total binary mass was deduced from the inspiral signal (see discussion in Sect.~\ref{sec:mass}). The total binary mass determines which of the relations has 
to be employed to convert the measured peak frequency to a NS radius. Note that all data are plotted as a function of $R(1.6~M_\odot)$ and hence the fiducial mass is not optimized to yield the tightest relations for total binary masses of 2.4~$M_{\odot}$ or 3.0~$M_\odot$. Here we assume that the determination of the binary masses does not contribute a significant error, but that the binary masses are known with sufficient precision as argued in Sect.~\ref{sec:mass}. We also point out that very similar relations can be constructed for unequal-mass binaries with a fixed total mass and mass ratio.

We here do not further discuss asymmetric binaries in great detail, but refer for instance to~\cite{2012PhRvD..86f3001B}. There it is shown that the peak frequencies of asymmetric binaries deviate only somewhat from the ones of the symmetric binaries of the same total mass. This can be also seen in Fig.~\ref{fig:asym}, which shows the peak frequency as function of $R_{1.6}$ for symmetric and asymmetric mergers. The symmetric and asymmetric binary systems are chosen such that they have the same chirp mass of $M_\mathrm{chirp}=1.1665~M_\odot$ (see Sect.~\ref{sec:mass}), which will be measured very accurately from the GW inspiral signal. This chirp mass corresponds to a symmetric setup with two stars of 1.34~$M_\odot$ (black symbols\footnote{We did not simulate 1.34-1.34~$M_\odot$ mergers, but interpolated linearly the $f_\mathrm{peak}(M_\mathrm{tot})$ relation, which is given by the results from 1.2-1.2~$M_\odot$ binaries and 1.35-1.35~$M_\odot$ binaries. In this binary mass range a linear 
interpolation is a very good approximation for symmetric systems (see Fig.~1 in~\cite{2014PhRvD..90b3002B}).} in Fig.~\ref{fig:asym}), i.e. $M_\mathrm{tot}=2.68~M_\odot$ (see Eq.~\eqref{eq:chirp}). The asymmetric configurations with  $M_\mathrm{chirp}=1.1665~M_\odot$ shown as blue symbols in Fig.~\ref{fig:asym} are 1.2-1.5~$M_\odot$ mergers, which are arguable the most asymmetric systems expected in the binary population (see discussion in Sect.~\ref{sec:mass}). Less asymmetric binaries lead to smaller deviations from the peak frequency of the symmetric system. Importantly, the data in Fig.~\ref{fig:asym} illustrate that even for the unlikely case that no information on the binary mass ratio is available and only the chirp mass was measured (Sect.~\ref{sec:mass}), the dominant postmerger frequency $f_\mathrm{peak}$ still determines NS radii very accurately. The data points deviate by at most 258~meters from a fit (solid line), which is constructed from the data of the symmetric and asymmetric binaries.

The precise form of the fits of the $f_\mathrm{peak}-R$ data in Fig.~\ref{fig:fpeak} and the deviations from the fits of course depend slightly on the chosen sample of candidate EoSs, which should preferentially include all EoSs that are acceptable based on current knowledge. Moreover, uncertainties in the numerical or physical model may also lead to slight differences in the empirical relations, which, however, will diminish in the future by more detailed simulations. Also, it is possible to choose different functional forms which may to some small extent affect the accuracy of the empirical relations. It is worth noting that in contrast to~\cite{2014arXiv1412.3240T}, we do not see difficulties in fitting a function to their data for the dominant oscillation frequency. Using the listed frequencies\footnote{Note that the dominant oscillation frequency is called $f_2$ in~\cite{2014arXiv1412.3240T} as well as in some other references.} and computing the radii of 1.6~$M_\odot$ NSs for the employed piecewise 
polytropic EoSs,
 it is easily possible to fit for 
instance a bi-linear (or quadratic) 
function to the data for a fixed $M_\mathrm{tot}$. This, for instance, results in deviations of at most 214 meters for the 1.3-1.3~$M_\odot$ mergers, and thus confirms our conclusions that a measurement of $f_\mathrm{peak}$ is sufficient for a radius determination. 

In this context we note that the peak frequencies from our numerical simulations agree well with those of~\cite{2011PhRvL.107e1102S,2013PhRvD..88d4026H,2014PhRvL.113i1104T,2014arXiv1412.3240T,2015arXiv151006398F}. Hence, it would be surprising if the data of~\cite{2014PhRvL.113i1104T,2014arXiv1412.3240T} did not yield a close relation for a fixed binary mass while our data show a good correlation and the frequencies essentially agree.

\begin{figure}

\resizebox{0.5\textwidth}{!}{%
  \includegraphics{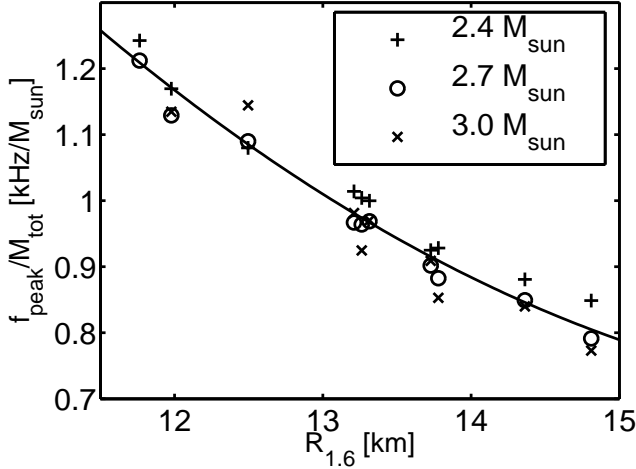}
}
\caption{Rescaled dominant postmerger GW frequency $f_\mathrm{peak}/M_\mathrm{tot}$ as function of the radius $R_{1.6}$ of a nonrotating NS with a gravitational mass of 1.6~$M_{\odot}$ for different EoSs and different total binary mass (plus signs for 2.4~$M_{\odot}$, circles for 2.7~$M_{\odot}$, crosses for 3.0~$M_{\odot}$) and a mass ratio of unity. }
\label{fig:fpeakres}
\end{figure}

In~\cite{2012PhRvD..86f3001B} it was argued that relations between $f_\mathrm{peak}$ and NS radii are not unexpected because the dominant emission is generated by the fundamental quadrupolar oscillation mode~\cite{2011MNRAS.418..427S}, whose frequency is known to scale tightly with $\sqrt{M/R^3}$ for nonrotating NSs (see~\cite{1998MNRAS.299.1059A}). This scaling for nonrotating stars suggests that it may be possible to rescale the frequency by dividing by $\sqrt{M_\mathrm{tot}}$ to describe the data of different total binary masses by a single relation. While this is indeed possible, empirically we find that $f_\mathrm{peak}/M_\mathrm{tot}$ results in an even tighter relation. Figure~\ref{fig:fpeakres} shows the rescaled peak frequency $f_\mathrm{peak}/M_\mathrm{tot}$ as a function of $R_{1.6}=R(1.6~M_\odot)$ for total binary masses of 2.4~$M_{\odot}$, 2.7~$M_\odot$, and 3.0~$M_\odot$. The rescaled frequencies can be described by the quadradic least-squares fit
\begin{equation} \label{eq:uni}
f_\mathrm{peak}/M_\mathrm{tot}=0.0157 \cdot R_{1.6}^2   -0.5495 \cdot R_{1.6} +  5.5030,
\end{equation}
with $f_\mathrm{peak}$ in kHz, the binary mass in $M_\odot$ and $R_{1.6}$ in km. The data in Fig.~\ref{fig:fpeakres} deviate by at most 500 meters from the fit. The reason why $f_\mathrm{peak}/M_\mathrm{tot}$ yields a better universality than $f_\mathrm{peak}/\sqrt{M_\mathrm{tot}}$ (which results in deviations of up to $\sim 1$~km) may be that the radius is also mass dependent. A perfect scaling behavior with the mass cannot be expected considering the findings in~\cite{2014PhRvD..90b3002B}, which show that different EoSs can lead to different dependencies on the total binary mass (see Fig.~1 in~\cite{2014PhRvD..90b3002B}). While such a universal relation of the rescaled $f_\mathrm{peak}$ is theoretically interesting, from a practical point of view it is preferable to consider different sets of mass-dependent relations as in Fig.~\ref{fig:fpeak} (or possible interpolations between them) because they yield tighter relations, and the total binary mass will always be known from the GW inspiral signal. 
Concerning mass-independent relations for $f_\mathrm{peak}$ we also note that a relation between the dominant postmerger frequency and the tidal coupling constant, pointed out in~\cite{2015arXiv150401764B}, is interesting because it connects the inspiral and the postmerger signals. Such a relation is understandable because essentially both quantities are known to depend on the EoS and to scale with the NS radius. As in the case introduced above (Fig.~\ref{fig:fpeakres}) the practical use of such relations depends on their tightness and a spread of $\sim 500$~Hz may be large compared to a typical width of the main postmerger peak.

For a first assessment of the detectability of the dominant postmerger GW frequency by a morphology-independent burst search data analysis, see~\cite{2014PhRvD..90f2004C}. A more detailed study employing a principal component analysis and evaluating the detectability for discussed future detectors is presented in~\cite{Clark2015}.

\section{Gravitational collapse and estimates of the maximum mass} \label{sec:col}
Apart from determining NS radii as discussed in the previous section, the postmerger phase also offers the possibility to constrain the maximum mass of nonrotating NSs. The maximum mass is a key parameter regarding the properties of high-density matter because its value is determined by the very high-density regime that is usually not encountered in most observed NSs, which have a lower mass. The maximum mass is thus particularly interesting for probing exotic phases of matter, which potentially occur at higher densities. Here we outline three different methods to determine or at least constrain the maximum mass of nonrotating NSs. An important quantity in the following discussion is the threshold binary mass to prompt black hole (BH) collapse. If one considers binary mergers with different total binary masses for a given EoS, one finds that for a binary mass above a certain threshold the merging leads directly to the formation of a BH and no (transiently stable) NS remnant is formed because the merged 
object is unstable against 
gravitational collapse. Note that the gravitational collapse in a merger remnant sets in at densities below the maximum density of non-rotating NSs.

\begin{figure}

\resizebox{0.5\textwidth}{!}{%
  \includegraphics{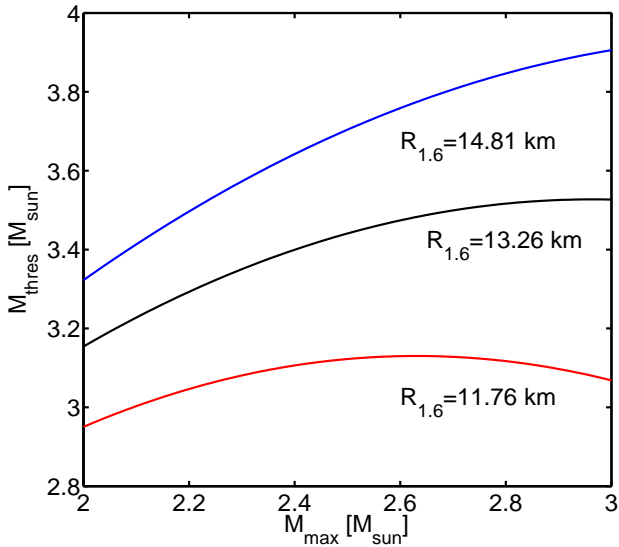}
}
\caption{Threshold binary mass for prompt BH formation as function of the maximum mass of nonrotating NSs for different $R_{1.6}$. The solid lines are given by Eq.~\eqref{eq:mthresc16}, which describes the empirical data with a good accuracy~\cite{2013PhRvL.111m1101B}.}
\label{fig:mththeo}
\end{figure}
The connection between this threshold mass and the maximum mass of nonrotating NSs has the following background. The maximum NS mass $M_\mathrm{max}$ is the threshold for BH formation 
of static, nonrotating stars, while the threshold mass $M_\mathrm{thres}$ represents the threshold for BH formation of differentially rotating, hot NSs. Thus, a quantitive relation between those two quantities is expected; and a determination of $M_\mathrm{max}$ is possible since the threshold mass can be determined from the binary masses of different merger events considering their respective outcome.

\begin{figure}

\resizebox{0.5\textwidth}{!}{%
  \includegraphics{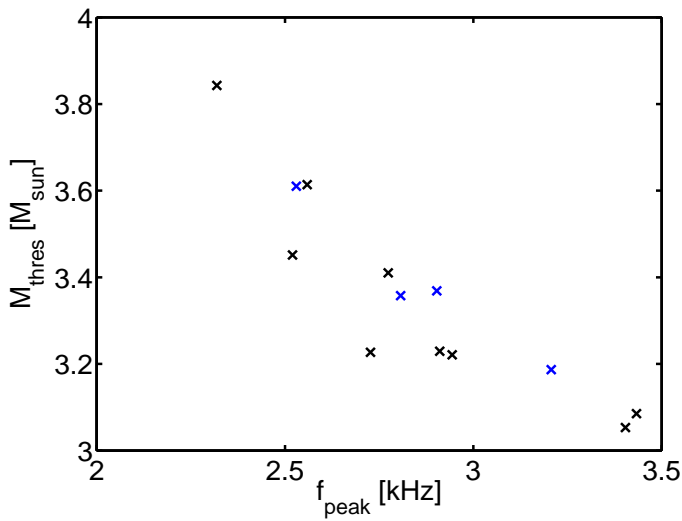}%
}
\caption{Threshold binary mass for prompt BH formation as function of the dominant postmerger GW frequency $f_\mathrm{peak}$ of 1.5-1.5~$M_\odot$ mergers for different EoSs.}
\label{fig:mthfpeak}
\end{figure}

{\em Method 1:} An analysis of a large number of simulations with different EoSs and different total binary masses has revealed that the threshold binary mass  $M_\mathrm{thres}$ depends in a particular way on the EoS~\cite{2013PhRvL.111m1101B}. This dependence can be described by TOV properties, which are uniquely defined by the EoS. With very good accuracy the threshold mass is given by
\begin{equation}\label{eq:mthresc16}
M_\mathrm{thres} = \left( -3.606 \frac{G M_\mathrm{max}}{c^2 R_{1.6}} + 2.38 \right) \cdot M_\mathrm{max}
\end{equation}
with the gravitational constant $G$ and the speed of light $c$. Similarly, it can also be expressed as
\begin{equation}\label{eq:mthrescmax}
M_\mathrm{thres} = \left( -3.38 \frac{G M_\mathrm{max}}{c^2 R_\mathrm{max}} + 2.43 \right) \cdot M_\mathrm{max},
\end{equation}
i.e. using the maximum mass and the compactness of the maximum-mass configuration. The coefficients in both equations are obtained by fits to the ratio $\frac{M_\mathrm{thres}}{M_\mathrm{max}}$, where $M_\mathrm{thres}$ is given by the results from simulations~\cite{2013PhRvL.111m1101B}. Equations~\eqref{eq:mthresc16} and~\eqref{eq:mthrescmax} reproduce the numerical results with a precision better than 0.1~$M_\odot$ (on average the deviations between the numerical and the estimated threshold mass are 0.024~$M_\odot$ for Eq.~\eqref{eq:mthresc16} and 0.031~$M_\odot$ for Eq.~\eqref{eq:mthrescmax}). Bear in mind that the numerical value of $M_\mathrm{thres}$ can only be determined with a finite precision because apart from uncertainties of the numerical and physical model only a limited set of models with different $M_\mathrm{tot}$ has been analyzed. \cite{2014arXiv1411.7975K} tested the collapse behavior for two EoS models, which were also employed in~\cite{2013PhRvL.111m1101B}, and found compatible results.
\begin{figure}

\resizebox{0.5\textwidth}{!}{%
  \includegraphics{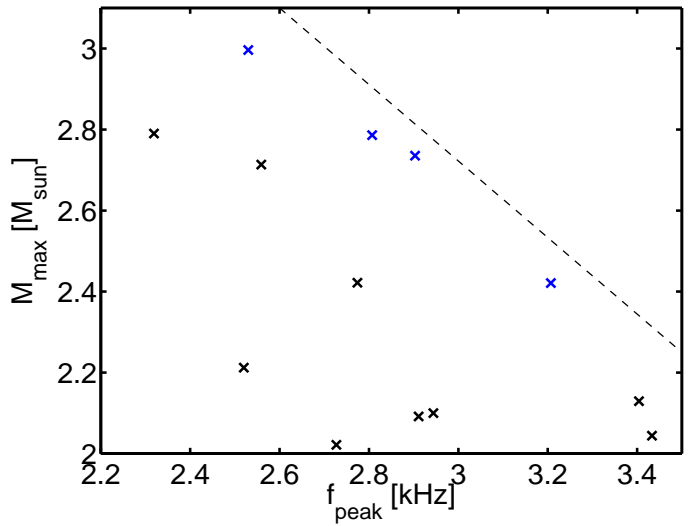}%
}
\caption{Maximum mass of nonrotating NSs as function of the dominant postmerger GW frequency $f_\mathrm{peak}$ for 1.5-1.5~$M_\odot$ mergers with different EoSs. The blue symbols mark EoSs which are particularly stiff at higher densities. The dashed line roughly indicates an upper limit on $M_\mathrm{max}$ for a given peak frequency, see Eq.~\eqref{eq:uplim}.}
\label{fig:mmaxfpeak}
\end{figure}

Using Eq.~\eqref{eq:mthresc16} Fig.~\ref{fig:mththeo} visualizes how $M_\mathrm{thres}$ depends on the maximum mass of nonrotating NSs for different chosen values of $R_{1.6}$. If the radius $R_{1.6}$ has been determined observationally, for instance through the detection of the peak frequency of a 1.35-1.35~$M_\odot$ merger (see Sect.~\ref{sec:rad}), then the threshold mass becomes an unambiguous function of $M_\mathrm{max}$. The threshold mass may be observationally determined by confirming or excluding the presence of postmerger GW emission of a NS remnant for several merger events with different total binary masses. (Note that the prompt formation of a BH leads to very weak postmerger GW emission, which can be distinguished from the formation of a NS remnant for near-by events and sufficient sensitivity.) The inversion of the relation between $M_\mathrm{thres}$ and $M_\mathrm{max}$ for a fixed $R_{1.6}$ can then be employed for an estimate of the maximum mass of nonrotating NSs.

We remark that a distinction between the prompt collapse and the formation of a NS remnant may also be possible by the observation of electromagnetic counterparts of mergers. In particular, the thermal emission of the ejecta, which are heated by radioactive decays, may be observable~\cite{1998ApJ...507L..59L,2005astro.ph.10256K,2010MNRAS.406.2650M,2013ApJ...767..124N}. The ejecta mass depends on the EoS, binary mass and mass ratio (see e.g.~\cite{2013ApJ...773...78B}). The prompt collapse to a BH leads to smaller amounts of unbound matter compared to mergers which form a NS remnant. This should lead to a noticeable difference in the light curves of these two different outcomes and may allow an observational identification of prompt collapse events. The detectability of electromagnetic counterparts to GW events has been addressed for instance in~\cite{2012ApJ...746...48M,2013ApJ...767..124N,2015MNRAS.446.1115M} and depends on (currently) uncertain details like the observer angle, the 
available instruments, the search strategy, the observing conditions, the ejecta mass, the outflow velocity, and the neutron-richness of the ejecta determining the opacity. The last three source properties are affected by the high-density EoS and the binary parameters.

{\em Method 2:} We point out that already a single GW event may provide a relatively precise estimate of the threshold mass and a constraint on the maximum mass of nonrotating NSs. Figure~\ref{fig:mthfpeak} displays the threshold mass as a function of the peak frequency of equal-mass mergers with $M_\mathrm{tot}=3.0~M_\odot$ for different EoSs. We use here the threshold masses estimated via Eq.~\eqref{eq:mthresc16}, which are very close to the numerical values. The blue symbols show the results for additional simulations with representative EoSs from~\cite{2010PhRvL.105p1102H}, which are derived within a chiral effective field theory at lower densities (up to roughly nuclear saturation density) and supplemented by extrapolations of the EoS at higher densities with physically motivated parameter variations (minimum NS mass of 2~$M_\odot$ and causality). For more information on this set of EoSs see~\cite{2012PhRvD..86f3001B}. Since these EoSs describe only cold NS matter, the EoSs are supplemented with an 
approximate 
treatment of thermal effects for the merger simulations (see e.g.~\cite{2010PhRvD..82h4043B} for details and an assessment of this approximation). In 
these calculations the ``ideal-gas'' index $\Gamma_\mathrm{th}$, 
which regulates the thermal pressure support, was chosen to be 5/3. See e.g.~\cite{2010PhRvD..82h4043B} and~\cite{2015arXiv150403982C} for a motivation of this value. One should keep in mind that the exact choice of $\Gamma_\mathrm{th}$ introduces an ambiguity and thus an uncertainty of the numerical values of certain quantities extracted from simulations. This is the reason why we did not include these models in the discussion in Sect.~\ref{sec:rad}. Note that from the original set of six EoSs in~\cite{2012PhRvD..86f3001B}, two models lead to the prompt formation of a BH for 1.5-1.5~$M_\odot$ binaries. This is fully compatible with the theoretically estimated threshold mass from the TOV properties of these EoSs via Eqs.~\eqref{eq:mthresc16} and~\eqref{eq:mthrescmax}.

Figure~\ref{fig:mthfpeak} shows that the determination of $f_\mathrm{peak}$ of a 1.5-1.5~$M_\odot$ merger yields the threshold mass with a precision of about $ 0.2~M_\odot$. Note that a total binary mass of 3.0~$M_\odot$ is only slightly larger than the masses of observed binaries for which precise mass measurements exist. Thus, a detection of a merger with $M_\mathrm{tot}=3.0~M_\odot$ may not be improbable. 

At first glance it may seem tempting to use the peak frequency of a 1.5-1.5~$M_\odot$ merger to fix $M_\mathrm{thres}$ (Fig.~\ref{fig:mthfpeak}), then to determine $R_{1.6}$ as discussed in Sect.~\ref{sec:rad} (Fig.~\ref{fig:fpeak}), and then to estimate $M_\mathrm{max}$ via Eq.~\eqref{eq:mthresc16}. This, however, represents an attempt to determine two independent quantities ($M_\mathrm{max}$ and $R_{1.6}$) by only one observable. In fact, a thorough error analysis for only one peak frequency measurement reveals that depending on the EoS the maximum mass can be determined only with a large error bar. (Notice the flat slope for small radii in Fig.~\ref{fig:mththeo}.) These considerations are summarized in Fig.~\ref{fig:mmaxfpeak}, which shows $M_\mathrm{max}$ directly as a function of the peak frequency of 1.5-1.5~$M_\odot$ mergers. Remarkably, $f_\mathrm{peak}(M_\mathrm{tot}=3.0~M_\odot)$ immediately yields an upper limit for $M_\mathrm{max}$, which for soft EoSs implies even a good estimate of $M_\mathrm{
max}
$ (recall the lower bound 
on $M_\mathrm{max}$ by the measurement of NSs with about 2~$M_\odot$). In Fig.~\ref{fig:mmaxfpeak}, the dashed line, which is given by
\begin{equation}\label{eq:uplim}
M_\mathrm{max, upper limit} = \frac{2}{3}f_\mathrm{peak} + 4.53,
\end{equation}
with the frequency in kHz and the mass in $M_\odot$, indicates an exclusion region and thus defines an upper limit on $M_\mathrm{max}$, which can be derived from a single detection of a merger with $M_\mathrm{tot}=3.0~M_\odot$. We stress that some of the EoSs within our sample (especially some of the models by~\cite{2010PhRvL.105p1102H}, blue symbols) have a speed of sound equal to the speed of light at higher densities. These models are thus maximally stiff, which is favorable for yielding high maximum masses. We thus conjecture that the dashed line represents a true upper limit and is not an artifact of the chosen sample of EoSs. We also refer to Fig.~16 in~\cite{2012PhRvD..86f3001B}, which shows a plot similar to Fig.~\ref{fig:mmaxfpeak} but for simulations of 1.35-1.35~$M_\odot$ mergers.

Fig.~\ref{fig:mmaxfpeak} shows that a single merger event can only constrain the maximum mass. However, we emphasize that the estimate of the threshold mass, e.g. through Fig.~\ref{fig:mthfpeak}, represents a viable goal on its own. The threshold mass is the crucial quantity to judge the outcome of a detected merger event. For instance, at larger distances only the chirp mass can be measured (which determines the total mass with some accuracy, see Sect.~\ref{sec:mass}). Having an estimate of $M_\mathrm{thres}$ available from a previous near-by event, the total mass of a distant merger may be crucial information for interpreting a possibly detected electromagnetic counterpart, for instance from thermal emission by the ejecta, which is powered by the radioactive decay of nucleosynthesis products~\cite{1998ApJ...507L..59L,2005astro.ph.10256K,2010MNRAS.406.2650M,2015MNRAS.446.1115M}. It is known that the prompt collapse leads to a reduced ejecta mass and thus to dimmer electromagnetic 
counterparts, which reach their peak emission on a shorter time scale e.g.~\cite{2013ApJ...773...78B}. Also, for the coincident detection of a short gamma-ray burst~\cite{1986ApJ...308L..43P,1989Natur.340..126E,2014ARA&A..52...43B} and a GW signal of a merger the information on the collapse behavior will provide valuable information and help to understand the conditions leading to this gamma-ray burst.

{\em Method 3:} While a single event can only yield an upper limit on the maximum mass $M_\mathrm{max}$, we have recently shown that several events with a measurement of the dominant postmerger GW frequency may suffice to determine the maximum mass~\cite{2014PhRvD..90b3002B}. Specifically, mergers with different binary masses (but in the most likely range of binary masses) can be employed for an $M_\mathrm{max}$ estimate. The procedure has the following background. Measuring the peak frequencies for two different total binary masses allows to estimate how the dominant GW frequency depends on $M_\mathrm{tot}$. This can be used to extrapolate $f_\mathrm{peak}(M_\mathrm{tot})$ and thus to estimate the behavior at higher binary masses, which is particularly sensitive to the maximum mass and the radius of the maximum-mass configuration. In essence, the absolute values of $f_\mathrm{peak}(M_\mathrm{tot})$ and its slope are very sensitive to $M_\mathrm{max}$ and $R_\mathrm{max}$ and thus yield an estimate of the 
maximum 
mass of 
nonrotating NSs 
with an accuarcy of roughly $\pm 0.1~M_\odot$. For details we refer to~\cite{2014PhRvD..90b3002B} and note that a slightly different approach is described in~\cite{BausweinWroclaw}, which results in a similar precision. We also point out that this extrapolation procedure determines the radius of the maximum-mass configuration and the maximum density of nonrotating NSs, which are quantities that are highly characteristic for the high-density regime of the EoS.

The advantage of this approach lies in the fact that it relies only on detections of mergers with binary masses in the most likely range, i.e. roughly between 2.4~$M_\odot$ and 3.0~$M_\odot$ (see e.g.~\cite{2012ARNPS..62..485L}). This contrasts the procedure to estimate $M_\mathrm{max}$ by a direct measurement of the threshold mass as sketched for Method 1. Estimating the threshold mass directly by determining the outcome of mergers with binary masses above and below $M_\mathrm{thres}$ requires several merger detections at higher binary masses, which are possibly very unlikely.

\section{Characterization of GW peak frequencies}\label{sec:sec}

\subsection{Origin of main and secondary GW peaks}
\begin{figure*}[!]

\resizebox{0.99\textwidth}{!}{%
  \includegraphics{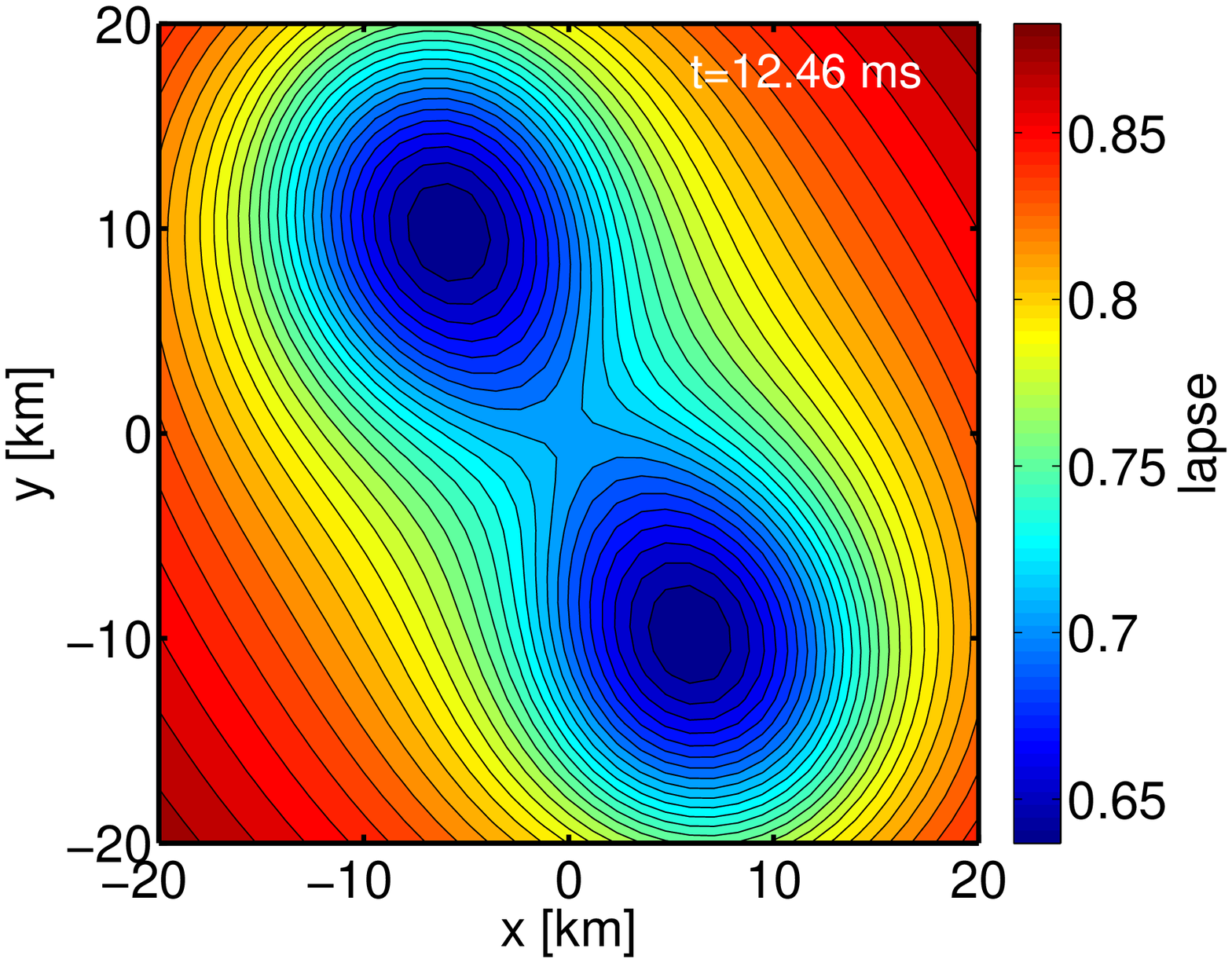}  \includegraphics{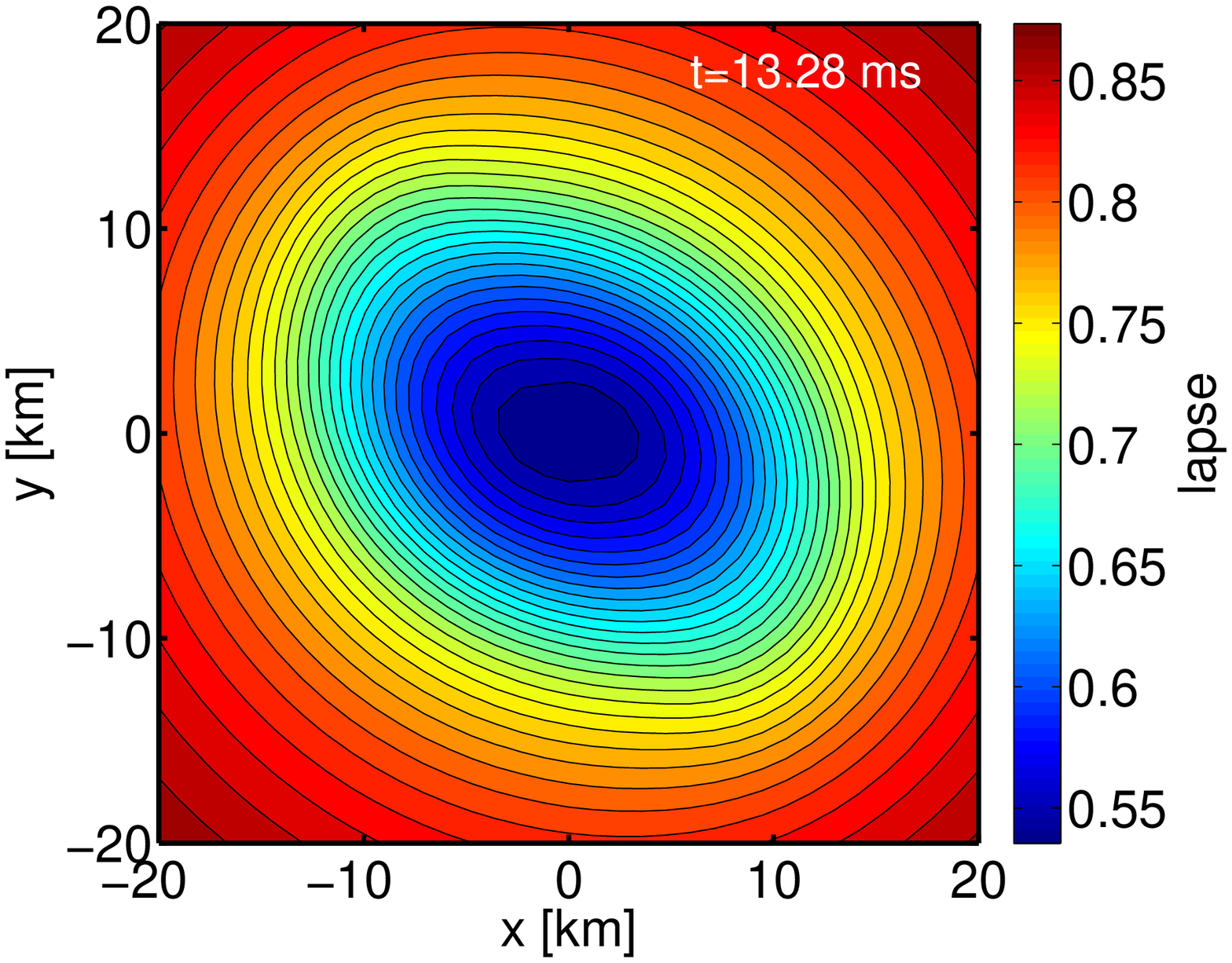}}
 \resizebox{0.99\textwidth}{!}{%
  \includegraphics{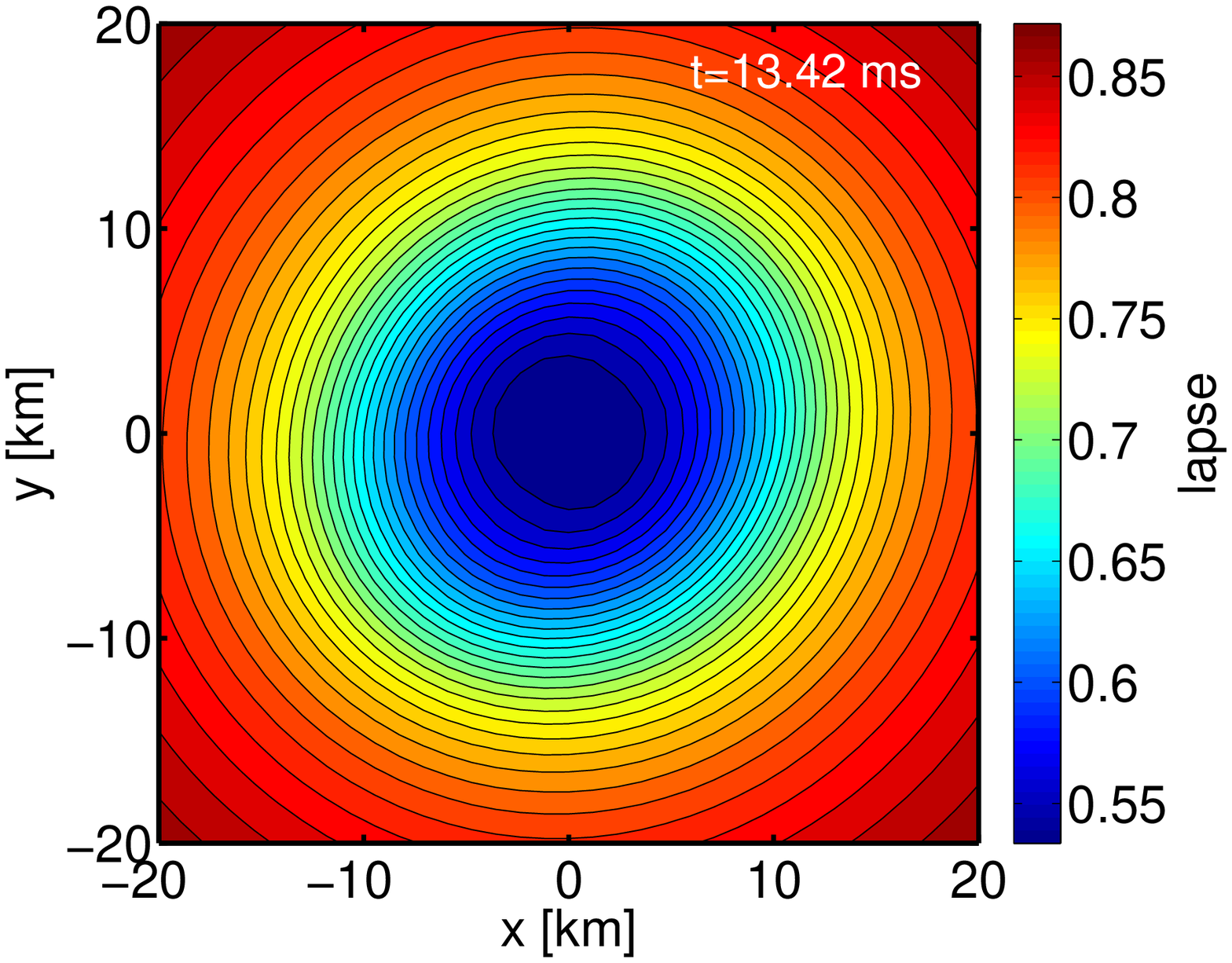}  \includegraphics{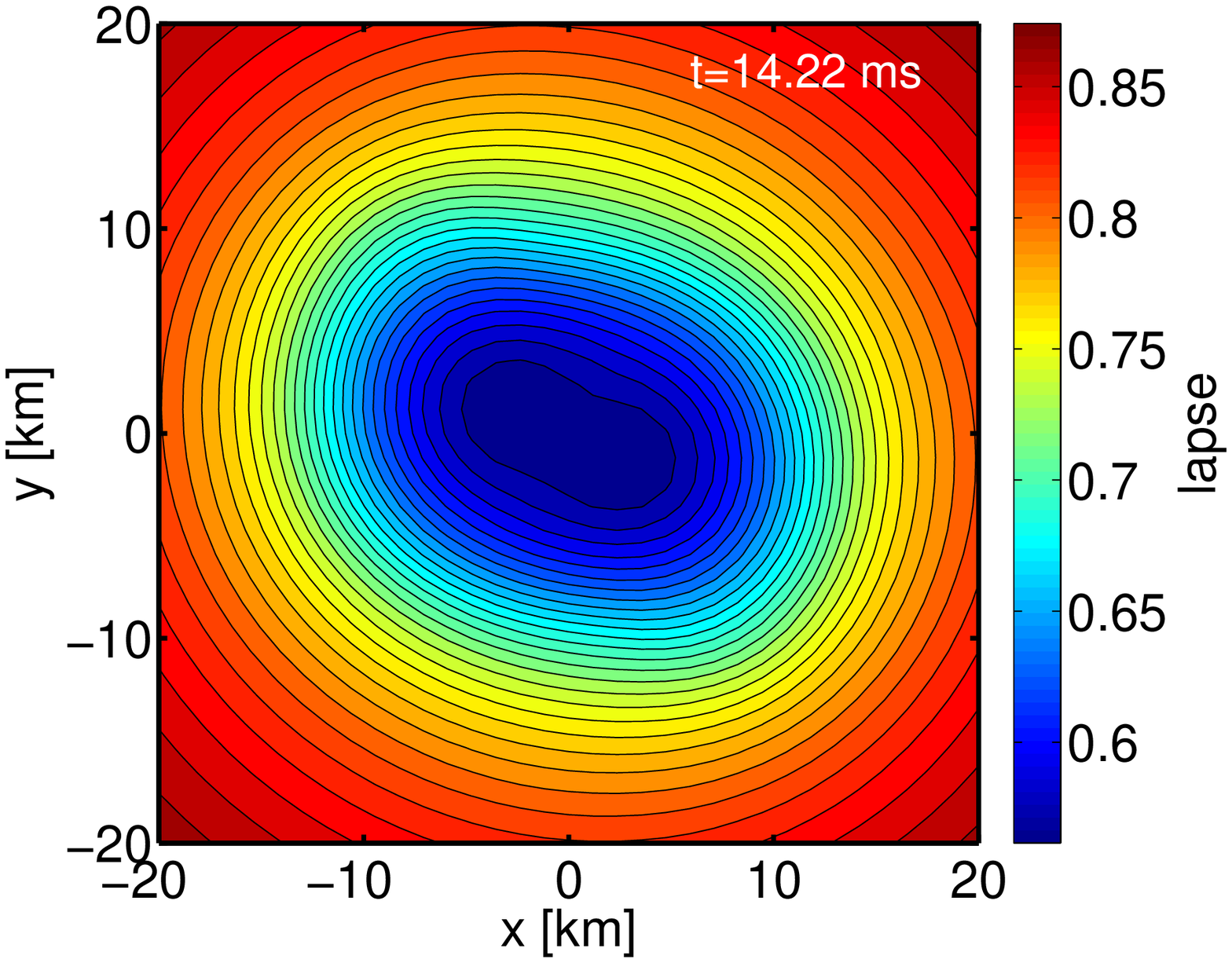}
}
\caption{Evolution of the lapse function of a 1.35-1.35~$M_\odot$ merger with the DD2 EoS in the equatorial plane. The snapshots should be compared to Fig.~2 in~\cite{2015arXiv150203176B} showing the density evolution for the same time steps of the same simulation.}
\label{fig:lapsesnap}
\end{figure*}
We start with a brief summary of our current understanding of the mechanisms which produce certain peaks in the GW spectrum~\cite{2015arXiv150203176B}. For the interpretation of the main and secondary peaks it is important to realize that the merger remnant can be considered as an isolated self-gravitating object~\cite{2011MNRAS.418..427S}. This view is supported by Fig.~\ref{fig:lapsesnap} showing the lapse function in the equatorial plane for a 1.35-1.35~$M_\odot$ merger with the DD2 EoS~\cite{2010NuPhA.837..210H,2010PhRvC..81a5803T}. The lapse function may be seen as the relativistic analog of the gravitational potential. In the upper left panel the system is still composed of two cores corresponding to the initial NSs. Shortly later, the two cores have merged into a single potential well. The time steps of the snapshots in Fig.~\ref{fig:lapsesnap} are the same as the ones in Fig.~2 of~\cite{2015arXiv150203176B}, which shows the 
evolution of the rest-mass density in the equatorial plane for the same model. It is remarkable that in the density evolution, a double-core structure persists for several milliseconds after merging, whereas the lapse function exhibits already a single core with partially strong deformations. (In the density plots corresponding to the upper right and lower panels of Fig.~\ref{fig:lapsesnap} a clear double-core structure is still visible.) These findings imply that the double-core structure visible in the density field for many milliseconds should not be interpreted as two gravitationally interacting objects, but rather as local overdensities moving in a single gravitational potential. Hence, the double cores should be viewed as tracers of the dynamics of a single, isolated object, rather than two independent dynamical objects.

\begin{figure}

\resizebox{0.5\textwidth}{!}{%
  \includegraphics{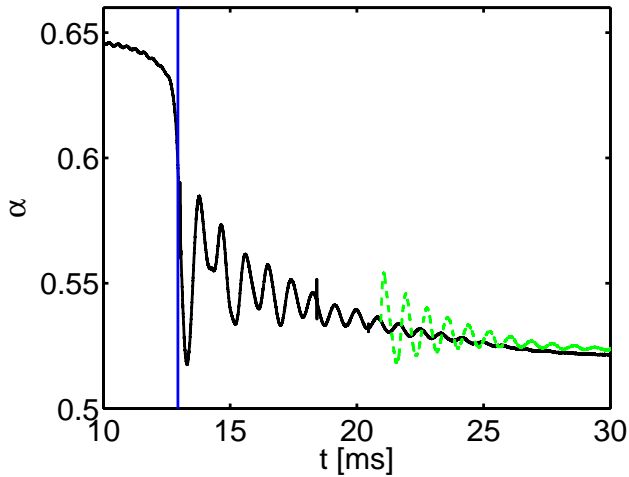}
}
\caption{Time evolution of the central lapse function of a 1.35-1.35~$M_\odot$ merger with the DD2 EoS (black line). The vertical line marks the time from which on only a single minimum of the lapse function is present. The dashed green line shows the evolution of the central lapse function of a late-time remnant of the same model where a quasi-radial velocity perturbation was added to excite the fundamental quasi-radial oscillation mode.}
\label{fig:lapseevol}
\end{figure}

\begin{figure}

\resizebox{0.5\textwidth}{!}{%
  \includegraphics{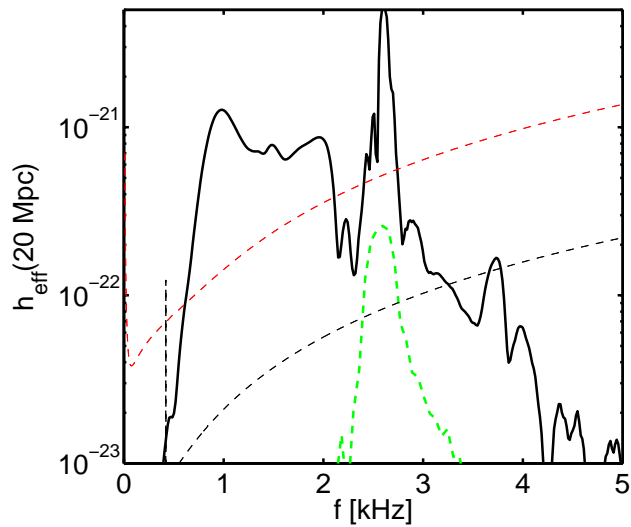}
}
\caption{GW spectrum of a 1.35-1.35~$M_\odot$ merger with the DD2 EoS (black line) given by $h_\mathrm{eff}=\tilde{h}(f)\cdot f$ with the Fourier transform of the waveform $h_{\times}$. The green dashed curve shows the GW spectrum of a simulation of a late-time merger remnant of the same model which was perturbed with a velocity field suitable to excite the fundamental quadrupolar fluid oscillation mode. Thin dashed lines show the anticipated unity SNR sensitivity curves of Advanced LIGO~\cite{2010CQGra..27h4006H} (red) and of the Einstein Telescope~\cite{2010CQGra..27a5003H} (black).}
\label{fig:fmode}
\end{figure}

The time evolution of the remnant structure is also summarized in Fig.~\ref{fig:lapseevol}. It shows the time evolution of the central lapse function, which may be interpreted as a measure for the compactness of the stellar object. The two initial NSs first touch at about $t=12.4$~ms. The vertical line indicates the time when the two cores in the lapse function (upper left panel in Fig.~\ref{fig:lapsesnap}) merge into a single core, which occurs already during the first compression phase, i.e. right during the final plunge. The subsequent oscillations in the central lapse function indicate quasi-radial oscillations (bounces and compressions of the remnant). The figure implies that matter accumulates into a single, gravitational trough already at a very early time in the remnant's evolution. Consequently, right from its formation the remnant's evolution can be described by the (non-linear) dynamics of a single, self-gravitating object. For instance, the dominant osillation can be associated with the 
fundamental 
quadrupolar fluid oscillation mode. This can be seen by adding 
a velocity perturbation to the remnant at late times, when it reaches a quasi-equilibrium and the GW emission essentially has diminished. We add an instantaneous perturbation of
\begin{equation}
\delta v_{\theta} = 0.4 \sin{\left(\pi \frac{r}{r_\mathrm{surface}(\theta)}\right)} \sin{(\theta)} \cos{(\theta)} \cos{(2\phi)},
\end{equation}
to the $\theta$-component of the coordinate velocity as a function the polar angle $\theta$, the azimuthal angle $\phi$ and the radial coordinate $r$. (Geometrical units are adopted.) The radial coordinate $r_\mathrm{surface}$ of the surface, which only depends on $\theta$, is defined by the coordinate at which the density drops below some threshold. We evolve the remnant with the added velocity perturbation for several milliseconds and extract the GW signal. The spectrum of the GW signal of the orginal simulation is compared to the spectrum of the perturbed model in Fig.~\ref{fig:fmode}. It is evident that the perturbation excites an oscillation mode with the same frequency as the dominant remnant oscillation, which strongly suggests that $f_\mathrm{peak}$ is the frequency of the $l=|m|=2$ fundamental mode\footnote{For this reason it was called $f_2$ in~\cite{2011MNRAS.418..427S} in the context of mode identification.}. This is corroborated by the extraction of the oscillation eigenfunction in~\cite{2011MNRAS.418..427S}, which shows a clean quadrupolar structure.

Also other (secondary) peaks in the GW spectrum can be explained by oscillation modes of the remnant. For instance, the peak at 1.5~kHz in Fig.~\ref{fig:spect} is produced by a quasi-linear interaction between the fundamental quadrupolar mode and the quasi-radial oscillation of the remnant~\cite{2011MNRAS.418..427S}. This can be shown by determining the frequency $f_0$ of the quasi-radial mode, which itself does not occur in the GW spectrum. However, it is very pronounced in the time evolution of the central lapse function (Fig.~\ref{fig:lapseevol}) and in other characteristic properties of the remnant, such as the maximum density and the size of the remnant. Again, by adding a suitable velocity perturbation to the late-time remnant, one can predominantly excite the quasi-radial mode and can extract its frequency from the time evolution of the central lapse function. This is visible in Fig.~\ref{fig:lapseevol}, where the green dashed curve shows the evolution of the radially perturbed model. The 
determination of 
the frequencies of the 
quadrupolar mode ($f_\mathrm{peak}$) and of the radial mode ($f_0$) reveals that a
secondary peak is expected to occur at $f_\mathrm{peak}-f_0$, which indeed is the case. The frequency coincidence confirms the nature of this secondary peak as being a coupling of two modes, which is why we refer to this feature as the $f_{2-0}$ peak. Another corresponding combination frequency can be recognized in Fig.~\ref{fig:spect} at approximately $f_\mathrm{peak}+f_0$. The peak at $f_\mathrm{peak}+f_0$ is observationally less interesting because of its weakness and the smaller sensitivity of GW detectors at higher frequencies, but it substantiates the importance of mode couplings in NS merger remnants.

Finally, there is one more secondary peak visible in the GW spectrum displayed in Fig.~\ref{fig:spect}. Recently, we provided evidence that this feature is generated by a spiral deformation which is created during merging. This deformation cannot follow the faster rotation of the inner remnant. The spiral deformation forms antipodal bulges, which orbit around the central part of the remnant for several milliseconds. Being a strong, non-axisymmetric, orbiting deformation, the antipodal bulges generate a GW signal at a frequency which is twice the orbital frequency of the bulges. A deeper analysis of the simulation data confirms this origin of the GW peak at 2~kHz for teh particular model (e.g. by extracting the orbital motion of the antipodal bulges and estimating their mass, by comparing the presence of the bulges and the presence of the secondary peak in the GW spectrum, and by computing GW spectra for the inner and outer parts of the remnant separately to estimate the contribution of the different remnant 
components 
to 
the different GW features). See~\cite{2015arXiv150203176B} for more details. Clearly, this so-called $f_\mathrm{spiral}$ feature cannot be explained within a perturbative approach. One can show by an analytic model that a peak with an appropriate strength in the GW spectrum can be produced by point particles of a few 0.1~$M_\odot$, which orbit for only a few milliseconds with an orbital separation 
of roughly the diameter of the inner remnant (see also Sect.~\ref{sec:toy}). 

\subsection{Classification of postmerger GW emission}
Considering models of NS mergers with varied binary masses and with different EoSs, one realizes that the three most prominent features in the GW spectrum can be explained by the three mechanisms detailed above: the fundamental quadrupolar mode, the coupling of the quadrupolar and the quasi-radial mode, and the orbital motion of antipodal bulges. The peak of the fundamental mode is present in all models and it is always the strongest feature. The presence and the strength of the different secondary features, however, are sensitively affected by the total binary mass. Depending on the total binary mass relative to the threshold mass $M_\mathrm{thres}$ (see Sect.~\ref{sec:col}) the different secondary peaks are more or less pronounced. This leads to a classification scheme that relies on the presence and strength of the different secondary features. One can identify three different types of NS mergers and corresponding GW spectra.

For relatively high $M_\mathrm{tot}$, i.e. close to but below $M_\mathrm{thres}$, the quasi-radial mode is strongly excited during merging, and consequently 
the $f_{2-0}$ feature is the most prominent secondary feature (Type I). In comparison, the $f_\mathrm{spiral}$ peak is weaker and may even be hardly visible in the spectrum. For moderate binary masses both secondary features are clearly visible and can be clearly distinguished in the spectrum (Type II). The secondary peaks have a roughly comparable strength. In a third type of GW spectra, the $f_{2-0}$ feature is absent (or hardly visible) and the GW peak by the spiral deformation is the most prominent secondary feature. This Type III occurs for relatively low total binary masses, that means for $M_\mathrm{tot}$ much below $M_\mathrm{thres}$. (An additional case is the prompt collapse to a BH as discussed in Sect.~\ref{sec:col} for $M_\mathrm{tot} \geq M_\mathrm{thres}$.)

Since the threshold mass depends on the EoS, the notation of a ``relatively high'' or ``relatively low'' binary mass is EoS-dependent as well. As a consequence, for $M_\mathrm{tot}=2.7~M_{\odot}$ all three types of mergers can occur depending on the EoS (see Fig.~5 in~\cite{2015arXiv150203176B}). For EoSs which lead to compact NSs, the initial stars merge with a higher impact velocity (see Fig.~3 in~\cite{2013ApJ...773...78B}), which is why the quasi-radial mode is strongly excited resulting in a pronounced $f_{2-0}$ peak (Type I). For stiff EoSs, i.e. less compact stars, the merging proceeds more gently with a lower impact velocity, which suppresses the strong excitation of the quasi-radial mode. Such models favor the formation of pronounced spiral deformations, and thus the $f_\mathrm{spiral}$ peak becomes particularly pronounced. The same reasoning explains the occurence of the different merger types for a fixed EoS dependending on the total binary mass.

We note that the classification of different merger simulations in~\cite{2015arXiv150203176B} is done based on the GW spectrum of the full signal (inspial and postmerger phase). Considering the postmerger spectrum only (i.e. windowing\footnote{By windowing we mean the application of a window function, e.g. a Tukey window, to the GW data, which can be used to exclude certain parts of the whole signal.} the signal appropriately) the two secondary peaks are weaker, which suggests that they interfere with power from the inspiral signal and that their actual strength is weaker. While the frequency of the inspiral signal may not even reach such high frequencies, some power may still be present at the frequencies of the secondary peaks because the inspiral signal is finite. We also note that this classification scheme, the description of the different mechanisms and the following discussion of the frequency dependencies hold for symmetric and mildy asymmetric binary mergers. The cases of strongly asymmetric mergers 
still need to be clarified as well as the nature of several other features which are even weaker than the secondary peaks. 

Finally, we mention that windowing the GW data has an impact in particular on the precise frequency of the $f_\mathrm{spiral}$ peak. Choosing the postmerger spectrum only or even excluding the very early postmerger phase, the $f_\mathrm{spiral}$ peak shifts to higher frequencies and its strength decreases. The latter is understandable because one excludes a part of the signal, when the mechanism generating the $f_\mathrm{spiral}$ feature is operating. The frequency shift may not be unexpected, given that the $f_\mathrm{spiral}$ peak is produced by a highly dynamical feature. For instance, one may expect that the antipodal bulges orbit at slightly smaller radii and thus lead to slightly higher orbital frequencies.

\subsection{Frequency dependence of secondary peaks (and EoS constraints)}
The dependence of the dominant oscillation frequency $f_\mathrm{peak}$ on the EoS was extensively discussed in Sect.~\ref{sec:rad}, and especially the potential for an accurate determination of NS radii (or, equivalently, NS compactnesses at fixed masses) was pointed out. The frequencies of the secondary peaks $f_\mathrm{spiral}$ and $f_{2-0}$ follow a behavior very similar to that of $f_\mathrm{peak}$, i.e. for a fixed total binary mass the frequencies are higher for EoSs which yield smaller NS radii. Relating the frequencies with the radii of nonrotating NSs (as in Sect.~\ref{sec:rad}) one finds similarly tight correlations for $f_\mathrm{spiral}$ and somewhat less tight for $f_{2-0}$ (see~\cite{2015arXiv150203176B}, see also~\cite{Clark2015} for a plot relating the secondary frequencies directly to the dominant frequency $f_\mathrm{peak}$). The frequencies follow the order $f_{2-0} < f_\mathrm{spiral} < f_\mathrm{peak}$.

In principle, secondary frequencies may be employed for EoS constraints~\cite{2014PhRvL.113i1104T,2014arXiv1412.3240T,2015arXiv150203176B}, however, a few issues are worth being mentioned. The secondary peaks are weaker than the main peak even if one takes into account the better sensitivity of GW detectors at lower frequencies. Hence, secondary frequencies will be more difficult to measure. In addition, the secondary peaks are broader in comparison to $f_\mathrm{peak}$ and often do not stand out clearly from the background. This will further impede an accurate determination of the secondary frequencies. Also, a detection of a single secondary peak will require additional information, e.g. from a measurement of $f_\mathrm{peak}$, to safely associate the secondary peak with either $f_\mathrm{spiral}$ or $f_{2-0}$. In comparison, a detection of the dominant peak frequency is more likely because of its strength, and, as discussed in Sect.~\ref{sec:rad}, a single detection of $f_\mathrm{peak}$ is sufficient to 
yield an accurate determination of NS radii.

Given the similarity to $f_\mathrm{peak}$ we do not further discuss the dependencies of the secondary peak frequencies here, but we refer to~\cite{2015arXiv150203176B,Clark2015} for more details. Regarding the implications for EoS constraints we, however, note that our simulations in~\cite{2015arXiv150203176B} do not confirm the existence of a mass-independent universal relation for the strongest secondary peak as claimed in~\cite{2014PhRvL.113i1104T,2014arXiv1412.3240T}\footnote{In~\cite{2014PhRvL.113i1104T,2014arXiv1412.3240T} no distinction between $f_\mathrm{spiral}$ and $f_{2-0}$ was made, but their secondary frequency $f_1$ should correspond to the strongest secondary peak, which for most cases may be the $f_\mathrm{spiral}$ feature.}. Instead, we find tight relations for the individual secondary frequencies for fixed total binary masses only. Note that there is no conflict between the simulation data of~\cite{2014PhRvL.113i1104T,2014arXiv1412.3240T} and~\cite{2015arXiv150203176B}, but different 
conclusions are explainable by the choice of the investigated binary setups. In fact, 
comparisons of the secondary and dominant frequencies for 
individual models yield a good quantitative agreement. 
The relation proposed in~\cite{2014PhRvL.113i1104T,2014arXiv1412.3240T} is built on simulations with a set of 6 EoSs, but with different binary-mass ranges for each EoS. The choice of range of the binary masses, however, affects the distribution of GW frequencies. This is the reason why an EoS-dependent choice of binary mass ranges introduces a bias (depending on whether the mass range for a given EoS is relatively high or low compared to the average), and why a universal relation does not exist if the same range of binary masses is chosen for all EoSs as in~\cite{2015arXiv150203176B}. Apart from this, for a robust universality and mass-independence the mass range should comprise approximately the range that is expected from observed binaries, and not only a small variation of 0.2~$M_\odot$ in $M_\mathrm{tot}$. Finally, we note that a set of only 6 EoSs may not be sufficient to allow for robust conclusions about the spread in empirical 
relations and thus the quality of certain relations and their usability for EoS constraints, because models which may possibly lead to outliers are not included. Since a mass-independent relation of the secondary frequencies does not exist, the relation proposed in~\cite{2014PhRvL.113i1104T,2014arXiv1412.3240T} cannot be employed for EoS constraints as proposed. However, as detailed in Sect.~\ref{sec:rad}, detecting the weaker secondary GW peaks is not essential for NS radius measurements if the stronger, dominant postmerger oscillation was measured.

\section{Analytic model for postmerger GW emission}\label{sec:toy}
The understanding of the most prominent GW emission mechanisms as detailed in Sect.~\ref{sec:sec} motivate us to set up an analytic model for the postmerger GW signal. This model may form the basis for GW templates to be used in matched filtering GW searches (see~\cite{2014PhRvD..90f2004C} for the potential of template-based searches in comparison to morphology-independent burst search algorithms). We report the model here by specifying the x-y component of the reduced quadrupole moment of the GW source, which can easily be interpreted as the cross polarization of the GW signal along the polar direction. The other components of the quadrupole moment can be deduced with the same set of parameters, from which the complete GW signal can be derived (i.e. both polarizations in all emission directions). For instance, the plus polarization can be obtained by adding an appropriate phase shift. In~\cite{2015arXiv150203176B} we identify three main mechanisms that produce the dominant postmerger GW emission. As also 
detailed in Sect.~\ref{sec:sec} this includes the dominant oscillation by the fundamental 
quadrupole mode, the coupling of this mode with the quasi-radial mode of the remnant and the transient emission from a spiral deformation. All 
three mechanisms 
can be 
modelled by individual sine functions 
with given initial amplitude, initial phases and exponential damping behavior. Hence, our analytic model reads
\begin{equation}\label{eq:toy}
\begin{split}
h_{\times}\propto Q_{xy} = &A_\mathrm{peak} \exp{(-(t-t_0)/\tau_\mathrm{peak})} \\
&\sin{(2\pi f_\mathrm{peak}(t-t_0)+\phi_\mathrm{peak})} \\
&+A_\mathrm{spiral} \exp{(-(t-t_0)/\tau_\mathrm{spiral})} \\
&\sin{(2\pi f_\mathrm{spiral}(t-t_0)+\phi_\mathrm{spiral})} \\
&+A_\mathrm{2-0} \exp{(-(t-t_0)/\tau_\mathrm{2-0})} \\
&\sin{(2\pi f_\mathrm{2-0}(t-t_0)+\phi_\mathrm{2-0})},
\end{split}
\end{equation}
for $t\geq t_0$ with a starting time $t_0$.  The particular advantage of our model (for instance in comparison to the fits to GW signals described by~\cite{2013PhRvD..88d4026H}) is that the parameters are motivated by the underlying physical mechanisms. Hence, it is not difficult to choose appropriate values for these parameters if one wants to reproduce a given GW signal, e.g. from a numerical simulation. The frequencies $f_\mathrm{peak}$, $f_\mathrm{spiral}$ and $f_{2-0}$ can be chosen as found in the spectra. The amplitudes and the damping time scales can be estimated from the time evolution of the GW signal and adjusted such that the correspoding peaks in the GW spectrum coincide with numerically obtained spectra. Generally, the amplitude and the damping timescale of the fundamental mode (``peak'') should be the largest. However, for the model 
discussed above (Fig.~\ref{fig:spect}), $A_\mathrm{spiral}$ may have a comparable strength but a shorter damping time scale than the contribution from the fundamental mode. As argued in~\cite{2015arXiv150203176B}, the antipodal bulges of the spiral deformation disappear after a few orbits, which suggests a damping time scale $\tau_\mathrm{spiral}$ of the order of a few milliseconds.

The physical background of the different parameters is also advantageous because the individual parameters can be constrained to certain ranges. For instance, the damping time scales $\tau_{2-0}$ and $\tau_\mathrm{spiral}$ are always smaller than $\tau_\mathrm{peak}$. The restricted range of certain parameters is an important property of our model because it signficantly reduces the parameter space and thus the computational costs in future applications as templates in matched filtering searches. The range and number of parameters (for instance the amplitudes and the frequencies) may be constrained if information about the total binary mass (or at least the chirp mass) is available from the GW inspiral signal. In this context, we point out that for a fixed total binary mass and a fixed mass ratio the three frequencies are highly correlated~\cite{2015arXiv150203176B,Clark2015}. For symmetric binaries with $M_\mathrm{tot}=2.7~M_\odot$, $f_\mathrm{spiral}$ and $f_{2-0}$ are tight functions of $f_\mathrm{
peak}$, which can be well 
approximated by 
\begin{equation}\label{eq:spiral}
f_\mathrm{spiral}=0.8058 f_\mathrm{peak}  -0.1895,
\end{equation}
and
\begin{equation}\label{eq:20}
f_\mathrm{2-0}=1.0024  f_\mathrm{peak} -1.0798,
\end{equation}
with the frequencies in kHz. Since these relations are accurate with deviations of typically only a few 10~Hz, $f_\mathrm{spiral}$ and $f_{2-0}$ can be essentially eliminated from our model, which reduces the dimensionality of the parameter space. If this precision is not sufficient for template searches one can at least define a very narrow range of frequencies around the estimates of Eqs.~\eqref{eq:spiral} and~\eqref{eq:20}. Further correlations and constraints of the different parameters will be explored in future work. For an application of this model in a template-based GW search, which we leave to future work, possible refinements of the damping behavior or small frequency drifts may be considered as well. Also a certain rise time of the signal may be implemented or the model may be directly connected to an inspiral signal (see~\cite{2015arXiv150401764B}). For strongly aymmetric binaries modifications may be necessary.
\begin{figure}

\resizebox{0.5\textwidth}{!}{%
  \includegraphics{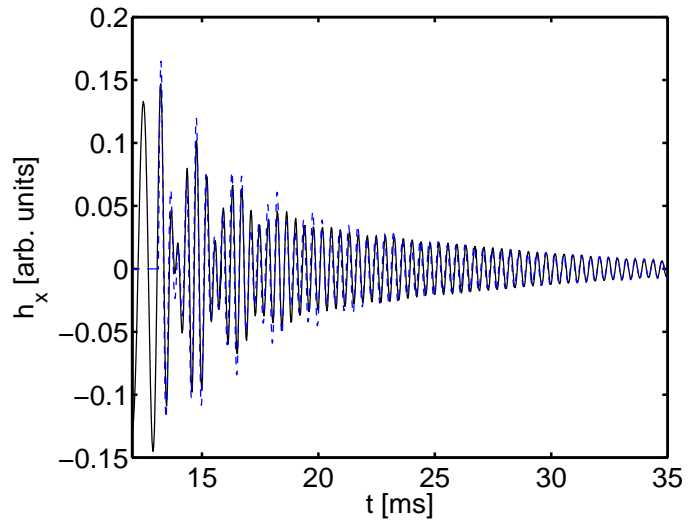}
}
\caption{GW signal of a 1.35-1.35~$M_\odot$ merger with the DD2 EoS (black). The analytic model GW signal is shown as dashed blue line (hardly visible since overlaid by the black curve) and is virtually indistinguishable from the actual signal.}
\label{fig:toysignal}
\end{figure}
\begin{figure}

\resizebox{0.5\textwidth}{!}{%
  \includegraphics{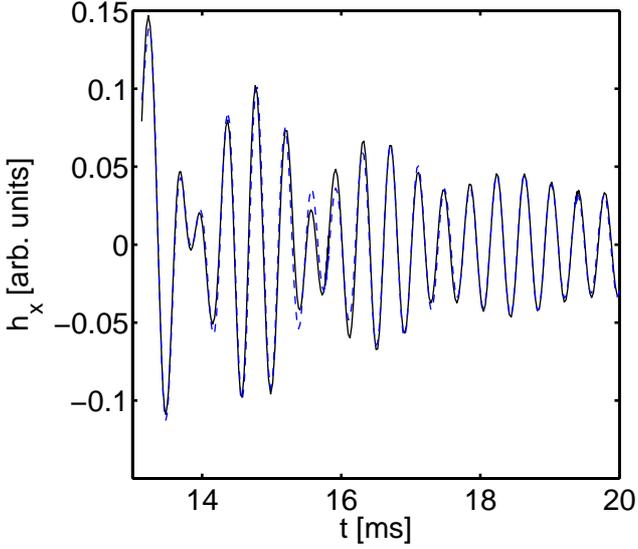}%
}
\caption{Initial phase of the postmerger GW signal of a 1.35-1.35~$M_\odot$ merger with the DD2 EoS (black). The analytic model GW signal is shown as dashed blue line.}
\label{fig:toysignal2}
\end{figure}

\begin{figure}

\resizebox{0.5\textwidth}{!}{%
  \includegraphics{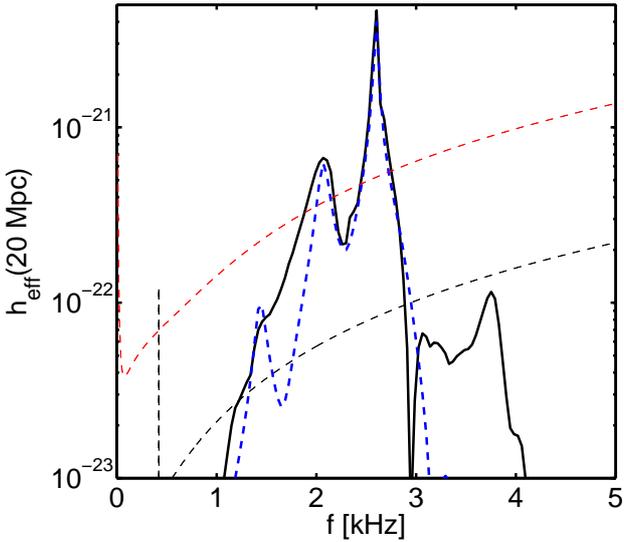}
}
\caption{Postmerger GW spectrum of a 1.35-1.35~$M_\odot$ merger with the DD2 EoS (black) given by $h_\mathrm{eff}=\tilde{h}(f)\cdot f$ with the Fourier transform of the waveform $h_{\times}$. The spectrum obtained from the analytic model in shown as dashed blue line. Thin dashed lines show the anticipated unity SNR sensitivity curves of Advanced LIGO~\cite{2010CQGra..27h4006H} (red) and of the Einstein Telescope~\cite{2010CQGra..27a5003H} (black).}
\label{fig:toyspec}
\end{figure}

We note that our model naturally leads to a time varying instantaneous GW frequency if the signal is interpreted as being produced by a single instantaneous frequency (see~\cite{2013PhRvD..88d4026H,2014arXiv1412.3240T,2014arXiv1411.7975K}). The interpretation of a single instantaneous frequency, however, cannot be supported by the underlying physical mechanisms producing the GW signal (see Sect.~\ref{sec:sec} and~\cite{2015arXiv150203176B,Clark2015}). Instead, there is evidence that several mechanisms with different frequencies contribute simultaneously to the signal. (See also the time-frequency map in~\cite{Clark2015}, which clearly shows distinct frequencies being simultaneously present and relatively stable in time.) This feature also distinguishes our analytic model from the model in~\cite{2014arXiv1412.3240T}, which describes only a single instantaneous orbital frequency which contributes to the GW signal and thus neglects the presence of the orbiting antipodal bulges generating the distinct $f_\mathrm{
spiral}$ 
peak.

To exemplify the potential of our analytic model we show in Fig.~\ref{fig:toysignal} the postmerger GW signal of a simulation with the DD2 EoS (black) together with the signal of our analytic model (dashed blue curve, Eq.~\eqref{eq:toy}) computed for a chosen set of parameters. As described above, the frequencies are taken from the GW spectrum and appropriate values are chosen for the remaining parameters. We stress that for this example the parameters were not determined by a fitting procedure but simply by a visual trial-and-error comparison of the numerical waveform and the model signal. The very good match between the waveforms in Fig.~\ref{fig:toysignal} is obtained without extensive fine-tuning of the parameters, but only by a crude, physically motivated choice of the parameters. An even better match can be achieved by an appropriate fitting procedure or more elaborate tuning of the parameters of the model (see~\cite{2013PhRvD..88d4026H}). This is shown in Fig.~\ref{fig:toysignal2}, where we employ a 
publicly 
available 
algorithm\footnote{We use the Covariance Matrix Adaptation Evolution Strategy downloaded from 
https://www.lri.fr/$\sim$hansen/cmaes\_inmatlab.html, which is described in~\cite{hansen2006eda}.} to generate an even better fit to the data. For clarity we show only the initial postmerger phase but note that the late phase is equally well reproduced as in Fig.~\ref{fig:toysignal}. In Fig.~\ref{fig:toyspec} we show the correspodning spectrum of the numerical waveform and of 
the analytic model, which apparently reproduces well the prominent features. Note that it is simple to include an additional feature in our analytic model to account for the peak at about 3.8~kHz, which is the coupling of the quadrupolar mode with the quasi-radial mode appearing at a frequency $f_{2+0}=f_\mathrm{peak}+f_0$. Since most of the parameters of this additional feature are already determined by the current model (Eq.~\eqref{eq:toy}), the consideration of this $f_{2+0}$ contribution would not significantly complicate the model. However, given the weakness of this peak and the lower sensitivity of GW detectors at higher frequencies we do not further investigate this option. Future work should investigate the performance of our model in fitting other waveforms, especially for asymmetric binary mergers, and the applicability of the model for template-based GW searches.

Finally, we note that for the $f_\mathrm{spiral}$ feature in the model shown in Figs.~\ref{fig:toysignal} and~\ref{fig:toyspec} we have chosen an amplitude of $A_\mathrm{spiral}=0.085$ and a damping time scale of $\tau_\mathrm{spiral}=3$~ms. An amplitude of this magnitude corresponds to two point particles of about 0.2~$M_\odot$ to 0.3~$M_\odot$ orbiting at roughly the surface of the remnant. This further substantiates the finding in~\cite{2015arXiv150203176B} that the $f_\mathrm{spiral}$ peak in the GW spectrum is generated by antipodal bulges which form during merging and which orbit around the inner remnant for only a few milliseconds.

\section{Impact of rotation}\label{sec:rot}
Most NS merger simulations have focussed on binaries with irrotational velocity profile, i.e. binaries with NSs that do not rotate intrinsically. This is justified by the conclusion that the viscosity is not sufficient to enforce co-rotation of the binary components by tidal interactions during the inspiral phase~\cite{1992ApJ...400..175B,1992ApJ...398..234K}. In addition, the intrinsic rotation of NSs in binaries is expected to be relatively slow since the stars cannot be spun up by accretion because of a missing donor star. In fact, the fastest known pulsar in a NS binary system has a rotation period of only 22~ms (see e.g. list in~\cite{2008LRR....11....8L}), which is slow compared to the orbital period of the binary prior to merging, which is of the order of roughly 2~ms. An intrinsic rotation period of 22~ms is also slow in the sense that the stellar structure is practically unaffected by the rotation, which only becomes important for spin periods below $\sim 5$~ms. Finally, one should bear in mind 
that 
most 
NSs in binaries rotate even more slowly than 22~ms, and that in the particular case of a NS spin of 22~ms the rotation will be further reduced by pulsar spin down due to magnetic dipole radiation until the binary components merge on a time scale of roughly 100~Myrs.

\begin{figure}

\resizebox{0.5\textwidth}{!}{%
  \includegraphics{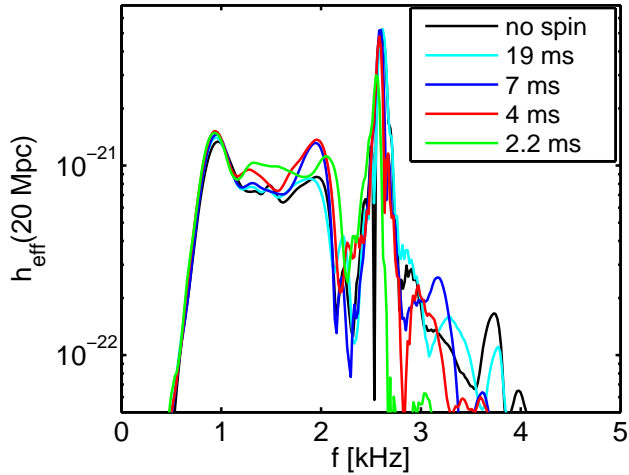}
}
\caption{GW spectra of a 1.35-1.35~$M_\odot$ merger with the DD2 EoS with different initial intrinsic NS spins.$h_\mathrm{eff}=\tilde{h}(f)\cdot f$ with the Fourier transform of the waveform $h_{\times}$.}
\label{fig:rot}
\end{figure}
Here we are mostly interested in the impact of intrinsic NS rotation on the GW signal, in particular on the dominant peak frequency of the GW spectrum, which can be employed for accurate NS radius constraints. With regard to the discussion in Sect.~\ref{sec:rad}, the most crucial question is whether intrinsic NS rotation can substantially alter $f_\mathrm{peak}$ for expected rotation rates compared to models without intrinsic rotation. (The relations like the ones in Fig.~\ref{fig:fpeak} could still be used to determine NS radii but with some loss of precision if the intrinsic rotation cannot be measured from the GW inspiral phase (but see e.g.~\cite{2013ApJ...766L..14H} for a discussion of spin measurements).) Comparing irrotational, co-rotating and counter-rotating spins (with respect to the orbital motion) only very little influence on $f_\mathrm{peak}$ was observed by~\cite{2007PhRvL..99l1102O}. More recent studies in full general relativity found a somewhat larger effect of the order of 100~Hz for 
intrinsic rotation rates roughly comparable to 22~ms~\cite{2014PhRvD..89j4021B}. A comparable influence was reported by~\cite{2014arXiv1411.7975K} but for roughly three times faster rotation rate, suggesting a somewhat weaker effect. However, these results and the observed impact of rotation may not be representative because the chosen binary systems are close to the threshold to prompt BH formation, and most of these simulated NS merger remnants experience the delayed collapse to a BH a few milliseconds after merging. For such binary setups the postmerger object undergoes strong changes in the stellar structure, and therefore the GW signal may be more sensitively affected by small changes, e.g. by a small difference in the angular momentum of the remnant. Moreover, the calculations of~\cite{2014PhRvD..89j4021B} employed an ideal-gas EoS, which may not represent a proper description of NS matter at all densities. \cite{2015arXiv150707100D} considered unequal-mass binaries with fast intrinsic rotation (
exceeding the rotation rate of 
the fastest known pulsar in binary systems) and found a shift of $f_\mathrm{peak}$ by only a few 10~Hz (in addition one setup with a misalignment between orbital and intrinsic rotation was investigated). 

In order to assess the impact of intrinsic rotation for a more representative setup we performed calculations of 1.35-1.35~$M_\odot$ binaries with the DD2 EoS and varied the initial intrinsic rotation rate using spin periods of 19~ms, 7~ms, 4~ms, and 2.2~ms. These simulations were then compared to the results from a calculation with irrotational velocity profile, which represents the usual setup used for deriving the results in Sects.~\ref{sec:rad} to~\ref{sec:sec}. Our simulations were performed with the code described in~\cite{2002PhRvD..65j3005O,2007A&A...467..395O,2010PhRvD..82h4043B,2012PhRvD..86f3001B}, which imposes the conformal flatness condition on the spatial metric for solving the Einstein equations~\cite{1980grg..conf...23I,1996PhRvD..54.1317W}. For simulations of irrotational binaries this approximation showed a very good quantitative agreement with fully general relativistic calculations (comparisons were reported in e.g.~\cite{2012PhRvL.108a1101B,2012PhRvD..86f3001B,2014PhRvL.113i1104T,
2015arXiv151006398F}).

Figure~\ref{fig:rot} demonstrates that the impact of initial NS rotation on the dominant postmerger GW frequency is practically negligible. Note that all tested models with NS spins rotate faster than the fastest known pulsar in a binary. For the model with the very fast rotation period of about 2.2~ms the peak frequency shifts by roughly 50~Hz relative to the model without initial NS spin, whereas the calculation with an initial NS spin period of 19~ms is practically identical to the computation without NS spin. One observes that models with faster initial spin lead to slightly lower peak frequencies and somewhat lower peak heights. This behavior is understandable from the fact that the remnants in such models have slightly more angular momentum. Hence, they are less compact and oscillate at lower frequencies. Also, the higher angular momentum slightly damps the excitation of the f-mode oscillation because of the centrifugal barrier, which is why the height of the main postmerger peak is slightly reduced. 
This can 
also be seen by considering the minimum of the central lapse function in the first compression phase (see Fig.~\ref{fig:lapseevol} for the model without intrinsic NS rotation), which reveals that mergers with initial spin are not as much compressed during the plunge and during the first compression phase as the corresponding model without spin.

Finally, the insensitivity of the peak frequency with respect to variations of the initial NS spin is understandable by considering the angular momentum of the remnant. Compared to the model without NS spin, the remnant's angular momentum is only slightly higher in models with spin because most of the angular momentum originates from the orbital motion of the inspiralling NSs. The contribution from the NS spins is only at the level of a few per cent (only for the model with a spin period of 2.2~ms the additional angular momentum slightly exceeds 10 per cent of the orbital angular momentum). Hence, differences in the remnant's structure are marginal, which explains why the peak frequencies are hardly affected by initial rotation even if the NSs spin relatively fast in comparison to measured spins in NS binaries. Therefore, the consideration of intrinsic NS spins can be safely 
neglected for the purpose of measuring the NS radii via the dominant postmerger oscillation frequency $f_\mathrm{peak}$ (Sect.~\ref{sec:rad}).

The different models displayed in Fig.~\ref{fig:rot} show that the secondary peaks may be affected by very fast intrinsic NS rotation. While the model with a rotation period of~19~ms does not exhibit important differences compared to the calculation with an irrotational velocity profile, the binaries with faster spinning NSs lead to a stronger $f_\mathrm{spiral}$ feature. This makes sense considering our explanations about the origin of this peak (Sect.~\ref{sec:sec}). Clearly, initial NS rotation favors the formation of antipodal bulges and thus leads to higher $f_\mathrm{spiral}$ peaks. Future work should also consider cases with different spin orientations and magnitudes of the individual binary components.

\section{Two-families scenario}\label{sec:two}
\begin{figure}
\resizebox{0.5\textwidth}{!}{%
\includegraphics{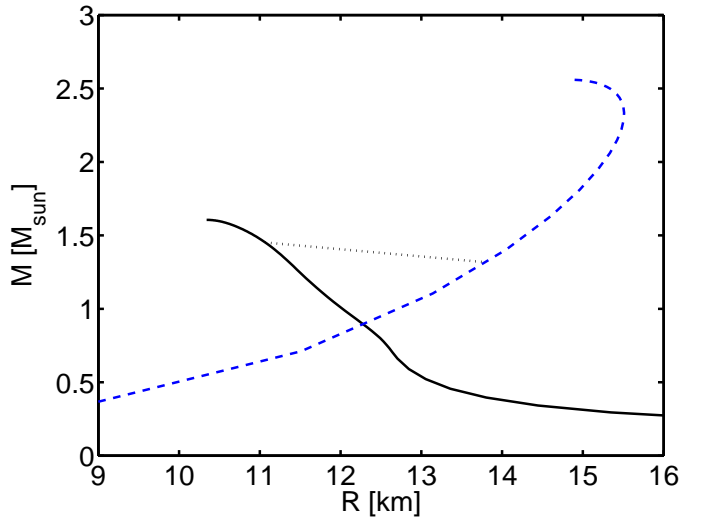}
  }
\caption{Mass-radius relation of a hadronic EoS (black curve) and an absolutely stable strange quark matter EoS (dashed blue curve) within the two-family scenario~\cite{2014PhRvD..89d3014D}. $M$ is the gravitational mass and $R$ the circumferential radius (EoSs provided by Giuseppe Pagliara). The dashed line indicates the conversion from a hadronic star to a quark star for a fixed rest mass.}
\label{fig:ssrm}       
\end{figure}

Recently, it has been proposed by~\cite{2014PhRvD..89d3014D} that two different families of compact stars, ordinary hadronic NSs and absolutely stable strange stars (SSs), might coexist under the assumption that the strange matter hypothesis is correct~\cite{1971PhRvD...4.1601B,1984PhRvD..30..272W} (see e.g.~\cite{2003ApJ...586.1250B,2004ApJ...614..314B,2015A&A...577A..40B} for similar ideas and~\cite{2015ApJ...810..134K} for arguments against absolutely stable strange matter). Specifically, the authors argued for a soft hadronic EoS and a stiff quark matter EoS. In this scenario, low-mass compact stars are NSs and compact stars with masses above a certain threshold are absolutely stable SSs, i.e. objects entirely composed of quark matter except for a dynamically unimportant crust of nucleonic matter with densities below the neutron drip density. A conversion of NSs to SSs may be triggered by different processes, e.g. mass accretion onto a NS. The mass-radius diagram in Fig.~\ref{fig:ssrm} 
illustrates such a jump from the NS branch to the SS branch. During the conversion (e.g.~\cite{1987PhLB..192...71O,1994PhRvD..50.6100L,2011PhRvD..84h3002H,2013PhRvD..87j3007P,2014PhRvD..89d3014D,2015arXiv150608337D}) the baryon number of the compact star is conserved, and the change in the binding 
energy leads to a different gravitational mass. While clearly 
speculative, the scenario 
of two families of compact stars represents a possible solution to the hyperon puzzle, i.e. the scenario is compatible with a hadronic EoS which is strongly softened by the occurrence of hyperons, and a maximum mass of compact stars above $~\sim 2~M_\odot$. Here we do not perform actual simulations which treat the conversion process and the detailed postmerger evolution, but based on theoretical arguments we discuss the observational signature of such a scenario without further commenting on the likelihood of the scenario or considering possible counter-arguments. For instance, it still remains to be explored to which extent the existence of SSs could lead to a cosmic flux of strangelets, which would prevent the existence of NSs because seed strangelets would convert a NS into a SS~\cite{1986ApJ...310..261A,1988PhRvL..61.2909M,1991PhLB..264..143C}. For SS EoSs which are compatible with the lower observational bound of the maximum mass of nonrotating compact stars, we expect that significant 
amounts of strange quark matter become gravitationally unbound during a merger of SSs~\cite{2009PhRvL.103a1101B}. Whether there exists a cosmic flux of strangelets, however, depends 
also on the fragmentation and survival probability of the unbound strange quark matter (see e.
g.~\cite{2015arXiv150403365H} for a discussion of a possible reconversion of quark matter nuggets to ordinary nucleonic matter). We focus on the GW emission of mergers of compact stars within the described scenario.

\begin{figure}

\resizebox{0.5\textwidth}{!}{%
  \includegraphics{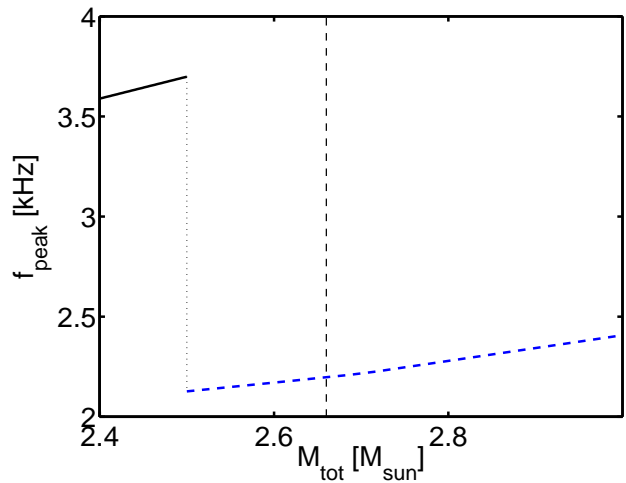}
}
\caption{Dominant postmerger GW frequency $f_\mathrm{peak}$ as a function of the total binary mass for symmetric mergers with a two-family scenario~\cite{2014PhRvD..89d3014D}. For low binary masses the merger remnant is composed of hadronuc matter (black curve), whereas higher binary masses lead to the formation of a strange matter remnant with a lower peak frequency (dashed blue curve). The vertical dashed line marks a lower limit on the binary mass which is expected to yield a remnant that is stable against gravitational collapse (see text).}
\label{fig:fpeaktwo}
\end{figure}

We consider only one exemplary model of the two-families scenario, i.e. one pair of nucleonic and quark matter EoSs (see Fig.~\ref{fig:ssrm}). (Specifically, for the hadronic EoS we use the relativistic mean field model of~\cite{2014PhRvC..90f5809D}, which includes hyperons and delta resonances and which is based on the model of~\cite{2012arXiv1207.2184S} with density-dependent coupling constants. We consider the model where the ratios between the coupling constants of the mesons to the $\Delta$ isobars and the coupling constants of the mesons to the nucleons are chosen to be $x_{\sigma \Delta}=1.15, x_{\omega \Delta}=1,x_{\rho \Delta}=1$. The EoS of absolutely stable strange quark matter adopts the model of~\cite{2014ApJ...781L..25F} with the parameter $X$ fixed to 3.5. The EoS tables were provided by G. Pagliara.) Clearly, variations to this model are possible, but the case discussed here is sufficient for the sake of outlining the special observational signatures of such a scenario and the 
qualitative behavior. We expect no qualitative 
differences for other 
realizations of the two-family scenario. We consider only the GW emission of the postmerger phase and assume that several GW observations of compact-star mergers are made with different total binary masses (see also~\cite{2010PhRvD..81b4012B} for an earlier discussion of the GW signature of SS mergers). We restrict the discussion to symmetric mergers.

For the particular model discussed here the transition of a NS to a SS occurs for NSs more massive than $\sim 1.45~M_\odot$ (gravitational mass), which corresponds to a central energy density of $\rho_\mathrm{conv}=1.5\times10^{15}~\mathrm{g/cm^3}$ (see Fig.~\ref{fig:ssrm}). This means that for total binary masses below 2.9~$M_\odot$ two NSs will merge, since the mass of each binary component is below 1.45~$M_\odot$. This should be considered as the generic case given that many observed NS binaries have a total mass below $\sim 2.9~M_\odot$. Prior to merging the maximum density in the initial stars starts to decrease because of the deformations in the very last inspiral phase. During merging the densities then increase again (see e.g. Fig.~3 in~\cite{2010PhRvD..81b4012B}). The maximum density in the merger remnant shows some oscillations and typically the maximum density in the remnant is larger than the one of the initial stars. However, in particular for low-mass mergers with $M_{\mathrm{tot}}$ below 
$\sim 2.4~M_\odot$, the 
maximum densities do not significantly exceed the one of the 
inspiralling stars because of the strong rotational (and thermal) effects in the remnant. Hence, for lower total binary masses the densities in the remnants remain below $\rho_\mathrm{conv}$ and are thus not sufficient to trigger the conversion of the NS matter to absolutely stable quark matter (here we neglect a possible influence of thermal effects on the conversion threshold). For larger $M_{\mathrm{tot}}$ the densities in the remnanst may exceed the central density of a nonrotating NS with $1.45~M_\odot$, and thus the conversion of NS matter to quark matter will set in. We note that a first density maximum is reached directly in the very first compression phase during merging. During the subsequent evolution of the remnant the densities are typically lower than this first maximum. Only after many ms, the redistribution of angular momentum in the remnant may lead to an increase of the maximum density above the first maximum. Hence, the conversion to absolutely stable strange quark matter is 
triggered 
right after merging or it occurs only on a 
longer time scale, after the remnant 
oscillations and thus the GW emission have ceased. In summary, this results in the following picture. Binaries below a certain binary mass threshold $M_{\mathrm{tot,conv}}$ lead to a merger remnant made of NS matter. For total binary masses above $M_{\mathrm{tot,conv}}$ the remnant encounters conditions that trigger the immediate conversion of NS matter to quark matter right during merging. Here we assume that the conversion takes place on a very short time scale that is shorter or at least comparable to the dynamical time scale of the remnant~\cite{2011PhRvD..84h3002H,2015arXiv150608337D}. Hence, a quark matter remnant forms essentially immediately, which has the same total rest mass (but a somewhat reduced gravitational mass compared to the corresponding NS merger remnant).

By performing a simulation for a 1.2-1.2~$M_\odot$ binary with the hadronic EoS, we find that the maximum density after merging is only somewhat lower than the transition density $\rho_\mathrm{conv}$, which triggers the conversion of a NS to a SS in a static configuration. From this we estimate that the binary mass which triggers a conversion is only slightly higher than 2.4~$M_\odot$, and thus we adopt in the following $M_{\mathrm{tot,conv}}=2.5~M_\odot$. We stress that thermal effects may actually trigger the conversion (during merging) already at lower densities. Hence, $M_{\mathrm{tot,conv}}$ may be somewhat lower, which we neglect for simplicity because it would require a detailed microphysical calculation of the temperature-dependent threshold.

Based on these remarks we can now estimate the postmerger GW emission of these binary mergers without actually performing simulations, which are technically non-trivial because the burning of NS matter to quark matter needs to be treated in an appropriate way including for instance an implementation of turbulent combustion. For the model discussed here we find $R_{1.6}=10.47$~km for the hadronic EoS and $R_{1.6}=14.53$~km for the SS EoSs by computing the TOV solutions (see Fig.~\ref{fig:ssrm}). Using these properties we can estimate the peak frequency of the postmerger phase by employing the fits discussed in Sect.~\ref{sec:rad}, which relate $R_{1.6}$ and $f_\mathrm{peak}$. From the fits we can compute the expected peak frequency for total binary masses of 2.4~$M_\odot$, 2.7 ~$M_\odot$ and 3.0~$M_\odot$ for both EoSs. With these data we can then interpolate linearly to predict the GW peak frequency as a function of $M_\mathrm{tot}$, where we distinguish the cases $M_\mathrm{tot}<M_{\mathrm{tot,conv}}$ 
and $M_\mathrm{tot}>M_{\mathrm{tot,conv}}$. In the former case we employ the estimates from 
the hadronic EoS, while for the latter case we 
use the peak frequency derived from $R_{1.6}=14.53$~km, i.e. the quark matter EoS. In Fig.~\ref{fig:fpeaktwo} one can read off which GW peak frequency is expected for a given total binary mass, which is measured by the inspiral GW signal (see Sect.~\ref{sec:mass}). It is clear that the formation of a SS remnant leads to dramatic changes of the postmerger GW signal compared to the GW emission from a NS remnant, which occurs for low total binary masses. Figure~\ref{fig:fpeaktwo} demonstrates that two families of compact stars result in a jump to lower peak frequencies at $M_{\mathrm{tot,conv}}$. Such a scenario can be clearly distinguished from only one family of compact stars, because in this case one expects a continuous increase of $f_\mathrm{peak}$ with $M_\mathrm{tot}$ (see Fig~1 in~\cite{2014PhRvD..90b3002B}). Observationally, this requires the detection of several merger events with different total binary masses to probe whether or not a jump to lower peak frequencies occurs with increasing binary mass. 

The vertical dashed line in Fig.~\ref{fig:fpeaktwo} displays the threshold mass for prompt BH formation for the hadronic EoS estimated via Eq.~\eqref{eq:mthrescmax}. Thus, at least up to this binary mass we expect the remnant to be stable independent of how fast the onset of the conversion to strange matter can stabilize the remnant.

We note that also SS EoSs roughly follow the $f_\mathrm{peak}-R_{1.6}$ relation, which justifies the use of the relations discussed above (see~\cite{2012PhRvD..86f3001B}\footnote{The SS EoSs considered in~\cite{2012PhRvD..86f3001B} describe bare SSs. While the inclusion of a nuclear crust would not have an impact on the peak frequency, which is mostly determined by the high-density regime of the EoS, the nuclear crust would lead to somewhat larger stellar radii (a few hundred meters) and thus would fulfill the $f_\mathrm{peak}-R_{1.6}$ relation even better than indicated in Fig.~11 of~\cite{2012PhRvD..86f3001B}.}). Given the small spread in the fits to the empirical data of the $f_\mathrm{peak}-R_{1.6}$ relations, the prediction of a frequency jump is a very robust feature of the two-family scenario. We remark that the transition from hadronic to quark matter leads to a change in the gravitational mass, which we do not take into account. Since these changes are small, they have only a secondary impact on the 
quantitative behavior in Fig.~\ref{fig:fpeaktwo}.

The results in Fig.~\ref{fig:fpeaktwo} represent the realization of only one possible model of the two-family scenario. However, we expect a qualitatively similar effect for other EoS models within the two-family scenario. In this context it is worth mentioning that for the particular model discussed here (i.e. the pair of hadronic and quark EoSs) the total binary mass $M_{\mathrm{tot,conv}}$ for formation of a SS remnant is relatively low compared to the expected mass range of NS binaries. As argued in Sect.~\ref{sec:mass}, binaries with total masses below 2.4~$M_\odot$ should be considered unlikely. Hence, the probability to detect a binary with $M_\mathrm{tot}<M_{\mathrm{tot,conv}}$ may not be high. However, we point out that there may be a second way to discern the two-family scenario from the ordinary model of only one family of compact stars, which may succeed with only one detection of a binary in the most likely mass range of about 2.7~$M_\odot$. GW detectors may measure finite-size effects during 
the 
last cycles of the inspiral~\cite{2010PhRvD..81l3016H,2010PhRvD..81h4016D,2012PhRvD..85l3007D,2013arXiv1310.8288F,2013arXiv1306.4065R,2013PhRvL.111g1101D,2014PhRvD..89j3012W,2015arXiv150305405A,2015PhRvD..91d3002L,2015arXiv150802062C}. Therefore, if the EoS information revealed from the inspiral phase is in stark conflict with the EoS constraint from the postmerger phase, this could be a strong indication for the two-family scenario, too. Note that for binaries in the most likely mass range, the inspiral probes only the hadronic EoS as long as $M_\mathrm{tot}<2\times 1.45~M_\odot$ for the model discussed here. Thus, the GW signal from the inspiral alone may not be sufficient to exclude the two-family scenario if no binary merger with $M_\mathrm{tot}>2\times 1.45~M_\odot$ is measured. Only the consideration of both merger phases may be conclusive in this respect. Finally, we note that also the possibility of SS-NS mergers for asymmetric binaries with individual masses above and below $M_{\mathrm{tot,conv}}/
2$ should be considered. We leave this for future work.

\section{Conclusions}\label{sec:sum}
We summarize the main ideas to infer EoS and NS properties from the postmerger GW emission of NS binaries and in particular we review the new findings of this work.
\begin{itemize}

\item The minimum requirement for accurate constraints of NS properties from the GW emission of NS merger remnants is the measurement of the total mass of the binary via the GW inspiral signal. The knowledge of the binary mass ratio is not critical, since the remnant's dominant oscillation frequency depends only weakly on the initial binary mass ratio. We illustrate that in the context of NS mergers the total binary mass can be estimated very well from the chirp mass, which will be measured with very high accuracy.

\item For a fixed total mass the dominant oscillation frequency of the NS merger remnant scales tightly with the radii of nonrotating NSs of  chosen fiducial mass for different EoSs. This in turn allows to determine the NS radius from a measurement of the dominant postmerger GW frequency, $f_\mathrm{peak}$. The error is given by the maximum deviation from the empirical relation between the GW frequency and the NS radius found among all EoSs. Choosing a fiducial NS mass somewhat higher than the mass of one of the merging NSs minimizes the deviations in the empirical relation and allows NS radius measurements with an accuracy of about 100 to 200 meters. Consequently, already a single detection of only the dominant peak frequency is already sufficient for a tight constraint on the NS radius and thus the high-density EoS through the frequency-radius relation.

\item We investigate asymmetric and symmetric binaries of the same chirp mass, which can be measured very accurately during the inspiral phase. The frequency-radius relation for the combined data from symmetric and asymmetric binaries is very tight (with maximum deviations below 300~meters). Therefore, even under the pessimistic assumption that only the chirp mass was measured accurately and no information on the binary mass ratio is available, the frequency-radius relation can be employed for accurate NS radius measurements.

\item For a broad range of binary masses (in the representative range between 2.4~$M_\odot$ and 3.0~$M_\odot$) a single relation between the dominant postmerger GW frequency and the NS radius can be constructed by rescaling the frequency with the total binary mass. Deviations from this relation are of the order of 500 meters. Thus, for practical purposes, relations for fixed total binary masses may be more useful because they are tighter and the total binary mass will be known. 

\item From our representative simulations we conclude that intrinsic rotation of the initial NSs has no significant impact on the dominant oscillation frequency for initial rotation rates as expected in NS binaries. Thus, intrinsic NS rotation can be safely neglected for deriving NS properties and EoS constraints from GW signals of the postmerger phase.

\item We present three different possibilities to constrain the maximum mass of nonrotating NSs from the collapse behavior of NS mergers or from the measurement of the dominant GW frequency. One of the key observations is that the binary threshold mass for prompt BH formation is well described by pure TOV properties, specifically by the maximum mass of nonrotating NSs and by the compactness of the maximum-mass TOV configuration. Thus, the threshold mass, which can be observationally determined, depends in a specific way on pure EoS properties. Also other properties of the maximum-mass configuration of nonrotating NSs like its radius or the maximum central density can be obtained from measuring postmerger GW emission, thus probing the very high density regime of NSs.

\item Our analysis provides further evidence that the dominant oscillation mode of the merger remnant is the fundamental quadrupolar fluid mode. In addition, our calculations show that the merger remnant forms a single, self-gravitating object right from the time of its formation. Although there are overdensities visible in the evolution of the density in the equatorial plane, which appear as a rotating double-core structure, these features should not be considered as independent dynamical features, but as tracers of the nonaxisymmetries of a single star.

\item Besides the dominant GW emission frequency (discussed above), we identify two distinct processes which produce prominent secondary peaks in the GW spectrum. One GW peak is prodcued by the orbital motion of an outer antipodal spiral pattern that forms during merging and persists for several milliseconds. Another GW peak is generated by the coupling of the fundamental quadrupolar mode with the quasi-radial mode. Depending on the exact system parameters (EoS and total binary mass) one or the other secondary feature is more prominent or, for some cases, both GW peaks can have comparable strengths. The presence and the prominence of the different features follows a clear behavior and can thus be embedded in a classification scheme of the postmerger dynamics and GW emission.

\item We present details of an analytic model for describing the structure of the postmerger GW signal that is based on the physical mechanisms producing the GW emission. A given postmerger GW signal can be well described by our analytic model. The model parameters are physically motivated and thus specific, physically motivated bounds on the parameters can be imposed, which is advantageous for a template-based GW search by reducing the computational effort. We point out the existence of certain parameter correlations which further reduce the dimensionality of the problem, and thus our description may form the basis for future matched-filtering GW data analysis.

\item We finally explore a more speculative scenario of two families of compact stars. Within this scenario massive compact stars are made of absolutely stable strange matter, whereas low-mass compact stars are ordinary NSs. Observationally this scenario can be revealed by a strong discontinuity in the dependence of the dominant postmerger GW frequency on the total binary mass. For the falsification or verification of this scenario the postmerger GW signal may be crucial, since the inspiral phase may only test the hadronic regime for the expected range of binary masses.
\end{itemize}

Overall, these conclusions show that the postmerger evolution is a highly interesting phase for understanding NS and EoS properties. In particular, it seems that certain characteristics are only accessible through the postmerger phase. Clearly, apart from advances in the modelling of the different aspects of this object, further studies of the data analysis capabilities of the advanced GW detectors are required. The highly rewarding prospects sketched in this work strongly motivate the construction of even more sensitive GW detectors, such as upgrades of Advanced LIGO and Advanced Virgo or the Einstein Telescope, e.g.~\cite{2010CQGra..27a5003H,Hild:2011np,2014RvMP...86..121A,2015PhRvD..91f2005M,Clark2015}.

\begin{acknowledgement}
We thank Matthias Hempel and Giuseppe Pagliara for providing EoS tables, and James Clark, Sebastiano Bernuzzi, Giuseppe Pagliara and Alessandro Drago for helpful discussions and suggestions.  A.B. is a Marie Curie Intra-European Fellow within the 7th European Community Framework Programme (IEF 331873). A.B. acknowledges support by the Klaus Tschira Foundation. Partial support came from “NewCompStar”, COST Action MP1304. The computations were performed at the Max Planck Computing and Data Facility (MPCDF), the Max Planck Institute for Astrophysics, and the Cyprus Institute under the LinkSCEEM/Cy-Tera project.
\end{acknowledgement}


 
%

\end{document}